\documentclass[12pt,preprint]{aastex}
\slugcomment{Draft}

\shorttitle{Electron Heating at SNR Shocks}
\shortauthors{Laming et al.}

\begin{document}

\title{Electron Heating, Magnetic Field Amplification, and Cosmic Ray Precursor Length
at Supernova Remnant Shocks}


\author{J. Martin Laming\altaffilmark1, Una Hwang\altaffilmark2,  Parviz Ghavamian\altaffilmark3 \& Cara Rakowski\altaffilmark4}


\altaffiltext{1}{Space Science Division, Naval Research Laboratory, Code 7684, Washington DC 20375
\email{laming@nrl.navy.mil}}
\altaffiltext{2}{University of Maryland\email{Una.Hwang-1@nasa.gov}}
\altaffiltext{3}{Dept. of Physics, Astronomy and Geosciences, Towson University, Towson, MD 21252
\email{pghavamian@towson.edu}}
\altaffiltext{4}{formerly of Naval Research Laboratory}

\begin{abstract}
We investigate the observability, by direct and indirect means, of a shock precursor arising from magnetic field amplification by cosmic rays.  We estimate the depth of such a precursor under conditions of nonresonant amplification, which can provide magnetic field strengths comparable to those inferred for supernova remnants.  Magnetic field generation occurs as the streaming cosmic rays induce a plasma return current, and may be quenched either by nonresonant or resonant channels.  In the case of nonresonant saturation, the cosmic rays become magnetized and amplification saturates at higher magnetic fields.  The precursor can extend out to $10^{17} - 10^{18}$ cm and is potentially detectable.  If resonant saturation occurs, the cosmic rays are scattered by turbulence and the precursor length will likely be much smaller.

The dependence of precursor length on shock velocity has implications for electron heating. In the case of resonant saturation, this dependence is similar to that in the more familiar resonantly generated shock precursor, which when expressed in terms of the cosmic ray diffusion coefficient $\varkappa $ and shock velocity $v_s$ is $\varkappa /v_s$. In the nonresonantly saturated case, the precursor length declines less quickly with increasing $v_s$. Where
precursor length proportional to $1/v_s$ gives constant electron heating, this increased precursor length could be expected to lead to higher electron temperatures for nonresonant amplification. This should be expected at faster supernova remnant shocks than studied by previous works. Existing results and new data analysis of SN 1006 and Cas A suggest some observational
support for this idea.

\end{abstract}

\section{Introduction}
Though diffusive shock acceleration has for many years been accepted as the theory of cosmic ray acceleration
at shock waves, the associated cosmic ray precursor has never been convincingly detected in an astrophysical
environment. The realization that cosmic rays in the preshock
region may amplify magnetic field by resonant \citep{bell01,lucek00} or
nonresonant \citep{bell04,bell05} instabilities, means that Alfv\'en Mach numbers and associated shock physics can be considerably different from the case without cosmic rays. These changes may
offer a means of detecting or inferring the existence of such a precursor, either by
direct imaging, or indirectly by the effects the precursor has on ambient plasma. Such plasma heating, produced by streaming cosmic rays, through their generation of plasma waves
which are subsequently damped by the ambient medium, has been recently considered in other
applications \citep[e.g.][]{nekrasov12,stroman12,wiener13}. Magnetic field amplification by similar processes also considerably increases the rate of cosmic ray acceleration, possibly allowing supernova remnants to generate galactic cosmic rays up to the ``knee'' (around $10^{15}$ eV) and maybe beyond in the cosmic ray energy distribution.

In related work, the quasi-thermal electron heating at a number of supernova remnant shocks has been investigated observationally. In shocks capable of accelerating
cosmic rays -- so-called collisionless shocks where the shock transition occurs on a length scale much shorter than the ion mean free path against Coulomb collisions -- the relative lack of collisions to enforce thermal equilibrium means that the electron and ion temperatures, $T_{e,i}$, are not necessarily equal. In general
$T_i > T_e$ following from the shock jump conditions applied separately to electron and ion fluids. The temperature ratio, $T_e/T_i$ has been shown to follow an approximate law
$T_e/T_i\propto 1/v_s^2$ law \citep[where $v_s$ is the shock velocity][]{ghavamian07,vanadelsberg08,ghavamian13}, at least for those cases where neutral H exists upstream of the shock to give rise to H$\alpha$ emission at the shock front itself. \citet{ghavamian07} suggested that such a behavior could arise if electrons are heated in a shock precursor established by the shock accelerated cosmic rays. Ths precursor extends a length $\varkappa /v_s$, allowing heating for a time $\varkappa /v_s^2$, where $\varkappa$ is the cosmic ray diffusion coefficient, assumed independent of the shock velocity
$v_s$. If the extent of the electron heating precursor is instead characterized by an ion gyroradius, $\sim v_s/\Omega$, where
$\Omega$ is the appropriate cyclotron frequency \citep[e.g.][]{cargill88}, $T_e/T_i$ independent of shock velocity would be expected, contrary to observations. No specific scenario of cosmic ray acceleration or magnetic
field amplification was discussed by \citet{ghavamian07}, though \citet{rakowski08} pointed out
that the waves envisaged by \citet{ghavamian07} are more efficiently excited at perpendicular
shocks, and that such a geometry would naturally arise as a result of cosmic ray driven magnetic
field amplification.

While magnetic field amplification is generally supported by Chandra observations of thin rims in X-ray synchrotron emission delineating the blast waves of SNRs,
the level and mechanism of saturation remain controversial. Observational estimates by various authors at Cas A \citep{vink03}, SN 1006 \citep{long03,yamazaki04} and Tycho \citep{warren05,chenai07} and other SNRs all summarized by \citet{volk05} generally reveal preshock magnetic fields in the range 100 - 500 $\mu$G, far greater than the typical
ambient magnetic field. Such fields can be consistent with the expectations of
\citet{bell04,bell05}, who suggested that the magnetic field amplifies through a nonresonant instability,
driven by the cosmic ray current, which saturates at a higher magnetic field level than the resonant instability. As this nonresonant instability proceeds, cylindrical cavities oriented along the pre-existing magnetic field develop. Here the high energy (i.e. unmagnetized) cosmic rays reside, with the background plasma, and presumably the magnetized cosmic rays expelled into the interfaces between neighboring cavities \citep[see e.g.][]{caprioli13}. Resonant magnetic field amplification produces much lower field, with $\delta B\sim B$, and cannot explain such observations. Other authors
\citep{luo09} argue that nonresonant magnetic field amplification must inevitably produce turbulence, and that the pitch angle scattering of cosmic rays in this turbulence reduces their anisotropy and velocity relative to the upstream medium, quenching the magnetic field amplification without producing any macroscopic structures upstream. Features of this scenario, especially the isotropization of the cosmic rays, are reproduced in various kinetic simulations  \citep{riquelme09,stroman09,gargate10}. This resonant saturation of the nonresonant instability
also produces lower magnetic field than the nonresonant saturation envisaged by \citet{bell04,bell05}, and again, lower field than is generally observed.

Yet another alternative means of magnetic field amplification at shock waves, that of the interaction of the shock with pre-existing turbulence, is discussed by \citet{giacalone07}. Solenoidal motions induced by this interaction generate magnetic field as it moves with the fluid. In its simplest form, this mechanism does not produce a shock precursor, and magnetic field is only amplified at the shock itself or just downstream from it, but can reach values as high as equipartition with the shocked plasma. \citet{beresnyak09} describe another version of such a model where the precursor in a cosmic ray modified shock interacts with turbulence and amplifies magnetic field upstream. Here the magnetic field amplification is driven by the cosmic ray pressure gradient, so the magnetic fields attained are
lower than in \citet{giacalone07}, but still higher than the nonresonant instability
of \citet{bell04} and \citet{bell05}.
Unlike the \citet{giacalone07} model, \citet{beresnyak09} allow the cosmic ray acceleration rate to be increased by the magnetic field amplification, with the proviso that the initial cosmic ray acceleration and modification of the shock must occur with an undisturbed upstream medium. \citet{drury12} revisit this and find somewhat lower
magnetic field amplification, but still sufficient to match observational inferences.

In this paper, we investigate how far the inference that $T_e/T_i\propto 1/v_s^2$ may hold
in conditions that cosmic rays nonresonantly amplify magnetic field, followed by either
resonant or nonresonant saturation. In section 2 we estimate the characteristic length of the shock
cosmic ray precursor under different conditions of magnetic field amplification and saturation,
to compare with $L=\varkappa /v_s$ in the purely resonant case. In section 3, we collect these
results and discuss how electron heating may vary with shock velocity
and degree of magnetization of the preshock cosmic rays. Section 4 applies these ideas to
observational results from the forward shocks of Cas A and SN 1006, and section 5 concludes.
Details of assumptions concerning the cosmic ray distribution function are given in an appendix.

\section{Magnetic Field Amplification}
\subsection{Introduction}
As mentioned above, preshock magnetic field may be amplified by either resonant or nonresonant
instabilities. Resonant amplification occurs when an individual cosmic ray gyrofrequency is Doppler shifted into
resonance with a specific wave mode (Alfv\'en or fast mode, both commonly referred to as Alfv\'en
waves when parallel propagating), and energy is transfered from particle to wave or vice versa occurs.
This typically saturates when $\delta B\sim B$ or less, because at this stage the wave-particle
resonance is lost and amplification ceases. Nonresonant magnetic field amplification is essentially
an MHD process. The current associated with the drifting cosmic rays (or more properly the return
current so induced in the background plasma) amplifies Alfv\'en waves, typically with $k_{\Vert}r_g >>1$ (in the resonant case $k_{\Vert}r_g\sim 1$, where $k_{\Vert}$ is the parallel wavevector, and
$r_g$ is the cosmic ray gyroradius). There is no individual wave-particle resonance.
\citet{bell04,bell05} has discussed nonlinear nonresonant saturation of this process, whereby unmagnetized
cosmic rays and their associated magnetic field expel the ambient plasma from cylindrical filaments, shutting down the return current as the plasma becomes magnetized. Other authors \citep[e.g.][]{luo09,gargate10} have discussed
a resonant means of saturating the nonresonant magnetic field growth, as the cosmic ray current is
limited by pitch angle angle scattering in turbulence. In the following subsections we estimate the
magnetic field amplification level and precursor distance over which this amplification occurs in
both regimes of saturation.

\subsection{Resonant Saturation of Nonresonantly Amplified
Field by Cosmic Ray Pitch Angle Diffusion}
Various kinetic simulations of nonresonantly amplified magnetic field \citep{riquelme09,stroman09,gargate10} suggest that cosmic ray induced magnetic field saturates not according to the \citet{bell04,bell05} scenario outlined above, but by the generation of turbulence that scatters the cosmic rays and reduces their drift relative to the upstream medium. We discuss here the level of magnetic field amplification and the length of the associated
precursor to be expected in such a scenario.

We follow in part the analytical description of this saturation process given by \citet{luo09}. The rate of change of the cosmic ray streaming speed ahead of the shock, $v_{CR}=\int fv_zd^3p$, where $f$ is the cosmic ray distribution function normalized to unity and $v_z$ the $z$-component of an individual cosmic ray velocity vector, due to scattering in turbulence is usually given as
\begin{equation}
{dv_{CR}\over dt}=\int f{dv_z\over dt}d^3p=-\int f\left<D\right>v_zd^3p,
\end{equation}
where $\left<D\right>=\int _{-1}^1\left(1-\mu ^2\right)Dd\mu /2$ with $D=\left(\pi/4\right)\left(\Omega _{i0}/\left<\gamma\right>\right)\left(\delta B^2/B^2\right)$, the cosmic ray pitch angle scattering diffusion coefficient.
Here $\Omega _{i0}$ is the proton cyclotron frequency calculated with the initial magnetic field
$B$, ($\Omega _i$ will be the same quantity calculated with the amplified field $\delta B$), $\left<\gamma\right>$ is the average cosmic ray Lorentz factor, and $\delta B^2/B^2 << 1$ is assumed. We extrapolate this to the case of magnetic field amplification by the nonresonant instability, where $\delta B^2/B^2 << 1$ no longer holds, as follows. Since the nonresonant instability preferentially grows magnetic field with
$kr_g>>1$, only cosmic rays with pitch angle cosines $\mu\simeq\pm 1/kr_g$ interact with the turbulence. The limits on the integral over $\mu$ become
$\pm 1/kr_g$ and $\left<D\right>=\left(\pi/4\right)\left(\Omega _{i0} /\left<\gamma\right>\right)\left(\delta B^2/B^2\right)/kr_g$. The particle scattering rate remains less than the gyrofrequency, even though $\delta B^2/B^2 >>1$, due to the presence of the factor $kr_g$ in the denominator. Assuming $\delta B\propto\exp\left(\Gamma t\right)$, we can integrate equation 1. \citet{luo09} insert a growth rate $\Gamma=\sqrt{2}k_{\Vert}v_A\sqrt{4\pi J_{CR}/k_{\Vert}cB -1}\simeq\left(6\eta kr_{g0}\right)^{1/2}\left(v_s/c\right)^{3/2}
\Omega _{CR}$ and derive $\delta B^2/B^2$ in the range 10-100. These values are lower than the
nonresonant saturation, and so saturation by resonant diffusion, suppressing the cosmic ray flow speed with respect to the upstream medium, will occur first if allowed.

The approximate expression for the growth rate above neglects the last term in the square root, and so overestimates the growth rate. This is not a very serious problem, except where the cosmic ray drift velocity is approaching the saturation value where $\Gamma\rightarrow 0$. We prefer to take $\Gamma =\sqrt{2}k_{\Vert max}v_A$, the maximum value of the growth rate at parallel wavevector $k_{\Vert max}=J_{CR}B/2\rho cv_A^2=2\pi J_{CR}B/c\delta B^2$, which may both be rewritten with $\omega _{pi}$ and $\Omega _i$ as the upstream ion plasma and cyclotron frequencies respectively as
\begin{eqnarray}
& & \Gamma ={3\eta\over\sqrt{2}}{\omega _{pi}\over\gamma _1\gamma _{max}}{\gamma _{max}-
3\gamma _1/4 \over\ln\gamma _{max}-1}{v_s^2v_{CR}\over c^3}{B\over\delta B}\nonumber\\
& & =0.013{\eta\sqrt{n_i}\over\gamma _1\gamma _{max}}{\gamma _{max}-3\gamma _1/4\over\ln\gamma _{max}-1}{B\over\delta B}
\left(v_s\over {\rm 5000~km~s}^{-1}\right)^2\left(v_{CR}\over {\rm 5000~km~s}^{-1}\right)~{\rm s}^{-1}
\end{eqnarray}
and
\begin{eqnarray}
& & k_{\Vert max}r_g={3\eta\over 2}{\omega _{pi}^2\over\Omega _i^2}{v_s^2v_{CR}\over c^3}
{\gamma _{max}-3\gamma _1/4\over\ln\gamma _{max}-1}{\gamma\over\gamma _{max}\gamma _1}\nonumber\\
& & =15340\eta n_i\left(v_s\over {\rm 5000~km~s}^{-1}\right)^2\left(v_{CR}\over {\rm 5000~km~s}^{-1}\right)\left(3\mu {\rm G}\over B\right)^2\left(B\over\delta B\right)^2
{\gamma _{max}-3\gamma _1/4\over\ln\gamma _{max}-1}{\gamma\over\gamma _{max}\gamma _1}.
\end{eqnarray}
Here, the current $J_{CR}$ is assumed to be carried by unmagnetized cosmic rays with Lorentz
factors $\gamma _1 < \gamma <\gamma _{max}$, as given in the Appendix, with an additional flux coming from cosmic rays escaping the shock upstream at $\gamma \ge\gamma _{max}$, given by
\citet{bell13} approximately as $n_{CR}v_s/4\gamma _{max}$.
The ratio of cosmic ray pressure to the shock ram pressure $\eta$ and $\gamma _1$ are assumed to be independent of $\delta B$. We substitute equation 3 into equation 1 and integrate the
right hand side over $p$, assuming $f\propto 1/p^4$. In equation 1, we rewrite $dv_{CR}/dt\rightarrow v_{CR}dv_{CR}/dz$. With $\delta B/B\sim\exp\left(
\Gamma z/v_{CR}\right)$, we put $dv_{CR}/dz\simeq -v_{CR}/z\simeq\Gamma/\ln\left(\delta B/B\right)$ in equation 1 to find
\begin{eqnarray}
& & \ln\left(\delta B\over B\right){\delta B^5\over B^5}\simeq {9\sqrt{2}\over\pi
}{\omega _{pi}^3\over \Omega _{i0}^3}{v_s^6\over c^6}\eta ^2\left(\gamma _{max}-3\gamma _1/4\over\ln\gamma _{max}-1\right)^2\left(1+\gamma _{max}\gamma _1\over\gamma _{max}^2\gamma _1^2\right)\nonumber\\
& & =8349\left(v_s\over {\rm 5000~km~s}^{-1}\right)^6\left(3\mu {\rm G}\over B\right)^3n_i^{3/2}\eta ^2\left(\gamma _{max}-3\gamma _1/4\over\ln\gamma _{max}-1\right)^2\left(1+\gamma _{max}\gamma _1\over\gamma _{max}^2\gamma _1^2\right).
\end{eqnarray}
This evaluates to $\delta B/B\sim 14\left(v_s/{\rm 5000~km~s}^{-1}\right)^{6/5}\left(\gamma _{max}/10^6\right)^{1/5}$ for $\eta \sim 0.1$, $n_i\sim 1$, and $\gamma _1 =1$, similarly to \citet{luo09}, but with
different dependencies of $\delta B/B$ on the shock velocity and other parameters. Comparing with \citet{gargate10} for example, where $\gamma _{max}\sim 10^3$, plugging in numbers for their models B1, B2 or B3 yields $\delta B/B\sim 14$, comparable to their simulation results.

This degree of magnetic field amplification is smaller than that inferred in the
observations cited above, though within uncertainties, once the magnetic field compression by
the shock is taken into account, the lower end of the postshock magnetic fields given above
approaching $\sim 100 \mu$G might be accessible.

The length scale over which this precursor develops is given by
\begin{equation}
L= {v_{CR}\over\Gamma}\ln\left(\delta B\over B\right)={\sqrt{2}\over 3}{c^3\over v_s^2}{\delta B\over B}\ln\left(\delta B\over B\right){\gamma _1\gamma _{max}\over\gamma _{max}-3\gamma _1/4}{\left(\ln\gamma _{max}-1\right)\over\eta\omega _{pi}}
\end{equation}
which evaluates to approximately $4\times 10^{10}\left(5000~{\rm km~s}^{-1}/v_s\right)^2\left(\delta B/B\right)\ln\left(\delta B/B\right)\gamma _1\ln\gamma _{max}/\eta\sqrt{n_i}$ cm for $\gamma _{max} >> \gamma _1$. With unmagnetized cosmic rays, $\gamma _1\simeq 1$, and $L\sim 10^{12}
\left(5000~{\rm km~s}^{-1}/v_s\right)^2\ln\gamma _{max}/\eta$ cm. The precursor to the shock in model B2 of \citet{gargate10} is thus predicted to be $8\times 10^{12}$ cm deep, in good agreement with the simulated value $\left(v_s/\Gamma _{max}\right)\ln\left(\delta B/B\right)\simeq 4\times 10^{12}$ cm.
Where $\delta B/B$ is approximately proportional to $v_s^{6/5}$, $L\propto v_s^{-4/5}\ln v_s$
which is close to the $L\propto 1/v_s$ discussed above.
Calculating $r_g$ with $\delta B$ rather than $B$ in equation 3 would give
$\delta B/B \propto v_s$ and $L\propto v_s^{-1}\ln v_s$, even closer to the behavior seen in the data. We emphasize that $L$ represents the cosmic ray precursor associated with magnetic field amplification, and is of course much smaller than the cosmic ray precursor associated with resonant wave generation and scattering, given approximately by $\left<D\right>/v_s$.

If the unmagnetized cosmic ray spectrum extends down to nonrelativistic energies so that $\gamma _1\sim 1$, then this precursor is predicted to be too small to be spatially resolvable, $\sim 10^{14}$ cm for $\eta\sim 0.1$ and $n_i\sim 1$. Significant magnetization of cosmic rays, increasing $\gamma _1$, lengthens the precursor, but also reduces the eventual magnetic field so long as resonant saturation remains effective. Such magnetization though is more likely in the case of nonresonant saturation discussed below, where higher magnetic fields can result.

\subsection{Nonresonant Saturation}
\citet{bell05} argues that cosmic ray induced magnetic field amplification should saturate as the unmagnetized cosmic rays and their associated magnetic field expel the ambient plasma from cylindrical filaments.

We can examine the expected magnitude of the amplified field at saturation by considering the nonresonant
instability \citep{bell04,bell05}.
At saturation, $1/r_g < k_{\Vert} < J_{CR}B/n_im_i cv_A^2=4\pi \left(n_{CR}^{\prime}+n_{CR}/4\gamma _{max}\right)qv_sB/c\delta B^2$ where $n_{CR}^{\prime}$ is the number density of unmagnetized cosmic rays with gyroradius $> r_g$. Consequently,
\begin{eqnarray}
& &\delta B <  \sqrt{4\pi n_{CR}^{\prime}v_sm_ic\left<\gamma ^{\prime}\right>}\nonumber\\
& & < \sqrt{12\pi\eta n_im_i v_s^3/c}\sqrt{\ln\gamma _{max}-\ln\gamma _1+1/4\over\ln\gamma _{max}- 1}\nonumber\\
& & < 5\times 10^{-4}\sqrt{\eta n_i}\left(v_s/{\rm 5000~km~s}^{-1}\right)^{3/2}
\sqrt{\ln\gamma _{max}-\ln\gamma _1+1/4\over\ln\gamma _{max}- 1} {\rm G.}
\end{eqnarray}
This represents an amplification
over the initial field, assumed to be 3$\mu$G, of a factor of 10 - 50 (taking $\eta \sim 0.1$,
$n_i\sim 1$ cm$^{-3}$), depending on the value assumed for the maximum Lorentz factor of magnetized cosmic rays, $\gamma _1$. This is about an order of magnitude higher than magnetic field amplification limited by resonant scattering.

We also calculate the magnetic precursor depth in the case that the growth is nonresonantly
saturated, according to \citet{bell04,bell05}. We assume ${\bf k}\Vert {\bf B}$, as seen
for example, even at oblique shocks, in simulations \citep{gargate12} and in {\it in situ}
observations \citep{bamert04}. Starting from Bell's expression for the growth rate
we write the time evolution of the amplified magnetic field $\delta B$
\begin{equation}
{d\delta B\over dt} = \sqrt{{J_{CR}Bk\over \rho c}-k^2v_A^2}\delta B =
\sqrt{{J_{CR}Bk\over \rho c}-{k^2\delta B^2\over 4\pi\rho}}\delta B.
\end{equation}
We integrate to find the time $t$ over which this magnetic field develops
\begin{equation}
t=\int _B^{\delta B}\sqrt{\rho c\over J_{CR}Bk}{1\over\sqrt{1-{kc\over 4\pi J_{CR}B}\delta B^2}}
{d\delta B\over\delta B}.
\end{equation}
We then put $\delta B = \sqrt{4\pi J_{CR}B/kc}\cos\theta$ to write
\begin{equation}
t=-\sqrt{\rho c\over J_{CR}Bk}\int _{\arccos\sqrt{kcB/4\pi J_{CR}}} ^{\arccos\sqrt{kc\delta B^2/
4\pi J_{CR}B}}\sec\theta d\theta = \sqrt{\rho c\over J_{CR}Bk}\left[\ln\left|\tan\theta +\sec\theta\right|\right]^{\arccos\sqrt{kcB/4\pi J_{CR}}} _{\arccos\sqrt{kc\delta B^2/
4\pi J_{CR}B}}.
\end{equation}
Evaluating,
\begin{eqnarray}
t=&\sqrt{\rho c\over J_{CR}Bk}\left[\ln\left|{\sqrt{1-kcB/4\pi J_{CR}}\over\sqrt{kcB/4\pi J_{CR}}}
+\sqrt{4\pi J_{CR}/kcB}\right|-\ln\sqrt{4\pi J_{CR}B/kc\delta B^2}\right]\cr
&=\sqrt{\rho c\over J_{CR}Bk}\ln\left|{\delta B\over B}\sqrt{1-kcB/4\pi J_{CR}} +{\delta B\over B}\right|.
\end{eqnarray}
At saturation, where $k=4\pi J_{CR}B/c\delta B^2$, with $\delta B>> B$
\begin{equation}
t=\sqrt{n_im_i\over 4\pi}{\delta B\over B}{c\over n_{CR}^{\prime}qv_{CR}+n_{CR}qv_s/4\gamma _{max}}\ln\left(2{\delta B\over B}\right)\simeq {c^3\gamma _1\gamma _{max}\over 3\eta v_s^2v_{CR}\omega _{pi}}{\ln\gamma _{max}-1\over\gamma _{max}-3\gamma _1/4}
{\delta B\over B}\ln\left(2{\delta B\over B}\right).
\end{equation}
The precursor depth is then
\begin{equation}
L={c^3\over 3\eta v_sv_{CR}\omega _{pi}}{\gamma _1\gamma _{max}\over\gamma _{max}-3\gamma _1/4}\left(\ln\gamma _{max}-1\right)
{\delta B\over B}\ln\left(2{\delta B\over B}\right)
\end{equation}
This is similar in expression to the case of resonant saturation of the nonresonant instability.
There is a missing factor of $\sqrt{2}$ in the numerical constant, and here $\delta B \propto v_s^{3/2}$. Also, for nonresonant saturation, $\gamma _1$ is likely to be significantly larger, yielding a longer cosmic ray
magnetic field amplification precursor, as well as a higher magnetic field than in resonant saturation. \citet{bell13} argue that $\gamma _1\sim\gamma _{max}$, in
which case $L\sim 10^{13}\gamma _{max}\ln\gamma _{max}/\eta\sqrt{n_i}$ cm.

We can estimate the physical size of the precursor by
taking $\gamma _{max}\sim 10^5$ \citep[e.g.][]{bell13,vink03}, to find $L\sim 10^{17}-10^{18}$ cm, which is potentially resolvable by e.g. the 0.5 arcsecond angular resolution of Chandra for relatively nearby galactic supernova remnants. \citet{morlino10} and \citet{winkler14} consider the case of SN 1006 specifically.
The different dependence of $\delta B$ on $v_s$ in this saturation case leads to
$L\propto v_s^{-1/2}\ln v_s$, assuming
$\gamma _1$ and $\gamma _{max}$ are independent of $v_s$. This is further from the $L\propto 1/v_s$ that
produces the constant electron heating with shock velocity, and suggests that one should look
at supernova remnants possibly subject to the Bell magnetic field amplification to see if evidence can be found for enhanced postshock electron temperatures because the magnetic precursor length
decreases less quickly with increasing shock speed.

Another means of nonresonant saturation is discussed by \citet{niemiec10}, following \citet{winske84}. The background plasma can be accelerated by the nonresonant mode, gradually
shutting down the cosmic ray current with respect to it. The
principal requirement for this is a cold cosmic ray ``beam'', where the perpendicular temperature is much lower than the parallel temperature, and when the beam is cold,
filamentation does not occur.  \citet{niemiec10} argue that such a case may occur
ahead of relativistic shocks, where only cosmic rays focussed along the shock velocity vector
are able to outrun the shock and contribute to magnetic field amplification. The application to
lower velocity SNR shocks appears less clear.

\subsection{Resonant or Nonresonant Saturation?}
Clearly, resonant saturation, if allowed, will restrict magnetic field amplification to only a
factor of a few over the preshock ambient field. However observations of SNRs generally reveal
much larger magnetic field amplification that this, suggesting that resonant saturation does not
occur. Either nonresonant magnetic field amplification proceeds until nonresonant saturation sets in, or some other mechanism of magnetic field amplification is at work.

Assuming that nonresonant field amplification is at work, what should determine whether resonant or nonresonant saturation will occur? We argue that when the parallel
wavevector of the amplification satisfies the inequality
\begin{equation}
{1\over r_g} < k_{\Vert} < {J_{CR}B\over\rho cv_A^2} = {\gamma n_{CR}\over n_i}{v_s^2\over v_A^2}
{\Omega _{CR0}\over v_s} < {\Omega _{CR0}\over v_s }{\delta B\over B}
\end{equation}
resonant scattering should begin to be inhibited. This is because when $k_{\Vert} < \Omega _{CR}/v_s =\left({\Omega _{CR0}/v_s }\right)\left({\delta B/B}\right)$ at a parallel shock,
the minimum drift velocity cosmic rays need to outrun the shock and form the precursor also moves them
out of resonance with parallel propagating waves.\footnote{At an oblique shock, the condition would be $k_{\Vert}\cos\theta _{bn} < \Omega _{CR}/v_s$ where
$\theta _{bn}$ is the shock obliquity, and the waves are assumed still parallel propagating.}
Substituting for $\delta B/B$ from equations 4 or 6 we get $v_s$ in the range
\begin{eqnarray}
& & \left(v_s\over 5000 {\rm km~s}^{-1}\right)< 0.21\left(B\over 3\mu {\rm G}\right)^{-13/14}n_i^{-1/4}
\eta ^{-3/14}\left(\ln\gamma _{max}-1\over\ln\gamma _{max}-\ln\gamma _1 +1/4\right)^{5/14} \nonumber\\ & &\rightarrow
0.64\left(B\over 3\mu {\rm G}\right)^{-6/5}n_i^{-1/5}
\eta ^{-1/5}\left(\ln\gamma _{max}-1\over\ln\gamma _{max}-\ln\gamma _1 +1/4\right)^{1/5},
\end{eqnarray}
respectively. To satisfy the conditions discussed by \citet{bell04,bell05}, we take $\gamma _1\rightarrow\gamma _{max}$, $\eta\sim 0.1$, and the upper limit on $\delta B$ derived from equation 6. The equation 14 shows that
resonant saturation is inhibited when
$v_s < \sim 10,000$ km s$^{-1}$ (dropping the explicit dependencies on other parameters).

In considering the competition between resonant and nonresonant magnetic field amplification,
\citet{marcowith10} argue for a gradual transition between the two
mechanisms over an order of magnitude in shock velocity, at similar values to those in
equation 14, with resonant amplification dominating at lower shock velocities, since at
saturation $\delta B\propto v_s$, whereas for nonresonant amplification $\delta B\propto v_s^{3/2}$. Conversely, we
have just shown that resonant saturation is favored over nonresonant saturation at higher shock velocities, although this depends on assumptions about the dependence of $\eta$ on $v_s$.

The arguments we have made above also tacitly assume a parallel shock geometry. \citet{bell05} gives the
angular dependence of the nonresonant growth rate, $\Gamma$. Keeping ${\bf k}\Vert {\bf B}$,
\begin{equation}
x^3 +x^2\left[2+\beta\right]+x\left[1+2\beta -{B^2j^2\over k^2v_A^4\rho ^2}\right]+\beta -{B^2j^2\over k^2v_A^4\rho ^2}\left[\left(\beta -1\right)\cos ^2\theta +1\right]=0
\end{equation}
where $x=\Gamma ^2/k^2v_A^2$, $\beta = c_s^2/v_A^2$, and $\theta $ is the angle between
${\bf k}$ and ${\bf j}$. At $\beta =1$, the explicit angular dependence disappears and
$\Gamma = kv_A\sqrt{Bj/k\rho v_A^2 -1}$. Putting $k=k_{\Vert max}$ from equation 3,
$Bj/k_{\Vert max}\rho v_A^2 =2$ and $\Gamma = k_{\Vert max}v_A$. For $0\le\beta\le 1$, $k_{\Vert max}v_A\le\Gamma\le 1.25k_{\Vert max}v_A$. We emphasize ``explicit'' angular dependence above,
because $k_{\Vert max}$ may depend on shock obliquity through its proportionality to $\eta = \left(\ln\gamma _{max}-1\right)c^2n_{CR}/3n_iv_s^2$. The variation of cosmic ray pressure $\eta$ with shock
obliquity is very uncertain. The threshold energy for injection into the diffusive shock
acceleration process increases with increasing obliquity \citep[e.g.][]{zank06}, reducing the
cosmic ray number density, but the acceleration rate in such geometry increases. The
generation of turbulence however is much more likely to decrease with increasing obliquity \citep[e.g.][]{laming13}, as does the cosmic ray current. These both make it likely that resonant scattering and saturation are even more inhibited in favor of nonresonant saturation in these cases of oblique shocks. Simply reinstating the obliquity in
equation 13 leads to $v_s$ gaining an extra factor $\cos ^{-5/14}\theta _{bn}\rightarrow\cos ^{-2/5}\theta _{bn} $ in equation 14.

In cases where cosmic ray acceleration is efficient \citet{reville13} show that amplified magnetic fields become
highly disorganized, and essentially isotropic. The initial shock obliquity has little affect on the final magnetic field. This might support some of our arguments about the transition to turbulence above, but by assuming that the cosmic ray acceleration is ``efficient'', \citet{reville13} avoid the shock injection issue. \citet{caprioli13} show that nonresonant saturation is still obtained at shock obliquities of 20$^{\circ}$, using large scale hybrid simulations, but they comment that the effect is particularly evident at parallel shocks. This implication is
contrary to our speculation above.

In summary, whether nonresonant magnetic field amplification saturates resonantly or nonresonantly depends
on the value of $\eta$ and its dependence on $v_s$, and remains unclear. Much of this uncertainty can be
traced to the problem of particle injection into diffusive shock acceleration, and so appears
unlikely to have a simple solution. However one clear difference between these two regimes appears
to be in the extent of the cosmic ray induced magnetic field shock precursor to be expected. This is something
that could be possibly exploited observationally to identify the saturation mechanism and we return to discuss it further below.

\subsection{Rakowski, Laming \& Ghavamian (2008) Revisited}
Following the inference of \citet{ghavamian07} that electrons at collisionless shocks may be heated in the cosmic ray precursor, \citet{rakowski08} estimated the
growth rate of lower hybrid waves in the cosmic ray amplified magnetic field. Lower hybrid waves are electrostatic ion oscillations directed almost perpendicularly to the magnetic field in conditions where the electrons are magnetized (electron gyroradius $<<$ wavelength), inhibiting the electron screening of the oscillation that would otherwise occur. The wave phase velocity perpendicular to the magnetic field is much smaller than that along it, allowing the wave to simultaneously be in resonance with unmagnetized ions and magnetized electrons.

\citet{rakowski08} calculated a kinetic growth rate of lower hybrid waves in a cosmic ray precursor, showing that the reactive growth for plausible cosmic ray distribution functions is zero. We correct here a small error in the final step of their calculation.
We give a revised version of their equation 5 for the growth rate of lower
hybrid waves in a cosmic ray precursor, where the cosmic rays are assumed to be in a ``kappa''-distribution with index $\kappa$, where $\kappa =2$ corresponds to the $p^{-4}$ spectrum familiar in first order Fermi acceleration,
\begin{eqnarray}
& & \gamma = \left( \frac{\pi}{\kappa}\right)^{3/2}\frac{q^{2}}{p_{t}k}
\frac{\omega^{3} n_{CR}}{\omega_{pi}^{2}+ \omega_{pe}^{2}\cos^{2}\theta}
\frac{(2\kappa-3)\Gamma(\kappa)}{\sqrt{2}\Gamma(\kappa -1/2)}
\int \delta(\omega - {\bf k} \cdot {\bf v}) \nonumber \\
& & \times \left\{ -\left[1+\frac{(p_{x}-mv_{s})^{2}}{2\kappa
p_{t}^{2}}\right]^{-\kappa} \frac{(p_{x}-mv_{s})}{p_{t}^{2}}{k\over\left|k\right|} +
\frac{\kappa}{\kappa -1}\frac{xv_{s}}{\varkappa ^{2}}\frac{\partial
\varkappa }{\partial p_{x}} \left[ 1 + \frac{(p_{x}-mv_{s})^{2}}{2\kappa
p_{t}^{2}} \right]^{1-\kappa} \right\} \nonumber\\
& & \times e^{-xv_{s}/D} dp_{x}.
\end{eqnarray}
This is modified from the original version by the factor $k/\left|k\right|$ multiplying the first term in curly
brackets. This arises from the scalar product ${\bf k}\cdot\partial f/\partial {\bf p}$, and restricts the region of $p_x$ where the growth rate may be positive to $0< p_x < mv_s$. When the momentum dependence of $\varkappa $ is neglected, maximum growth is found at $p_x=0.7 mv_s$ for $\kappa =2$ with rate
\begin{equation}
\gamma _1= 0.04 {n_{CR}\over n_i}\omega.
\end{equation}

It is not possible in this case for lower hybrid waves to stay in contact with the shock, as in \citet{laming01a} and \citet{laming01b}, but in the context of an extended cosmic ray precursor (as opposed to a narrow precursor due to shock reflected ions extending about one gyroradius upstream), this does not greatly affect their growth. If the second term involving $\partial \varkappa /\partial p_x$ dominates the growth rate, maximum growth is found at $p_x=\pm \sqrt{13}mv_s/2$, also for $\kappa =2$, with rate
\begin{equation}
\gamma _2= 0.17r {n_{CR}\over n_i}\omega,
\end{equation}
where the cosmic ray diffusion coefficient $\varkappa \propto p^r$.
At a parallel shock $r=1/3 - 1/2$ for Kolmogorov or Kraichnan turbulence respectively, whereas $r=1/9 - 1/6$ at a perpendicular shock \citep[see Appendix A in][]{rakowski08}. Thus the two terms are likely to contribute a similar order of magnitude to the growth rate. This is a factor of about 2 smaller than that originally given ($\gamma = 0.14 \left(n_{CR}/n_i\right)\omega$) in the most appropriate case of a perpendicular shock.
\citet{rakowski08} compared the growth rate for lower hybrid waves with that for magnetic field amplification, assuming that the same cosmic rays are responsible for both. They thus determine a critical Alfv\'en Mach number (of order 10) where magnetic field amplification ceases and lower hybrid wave generation takes over. This argument is most appropriate for resonant saturation, where the
magnetization of cosmic rays is not effective in saturating the growth. To explain other cases, the
growth rate of \citet{rakowski08} needs to be modified to include only the unmagnetized cosmic rays.

\subsection{Saturation of Magnetic Field Amplification by Electron Heating?}
To examine the effect of electron heating, we compare the growth rates for magnetic field amplification and lower hybrid waves.
At full nonresonant saturation, the unmagnetized cosmic rays and ambient plasma do not overlap, and the growth of lower hybrid waves should be suppressed. In such a case, the treatment of \citet{rakowski08} may become invalid.
However it is important to be aware that the inference $T_e/T_i\propto 1/v_s^2$ is determined from SNR shocks that have neutral material in their preshock media, and generally do not show strong X-ray synchrotron emission. As such these are unlikely to be the strongest cosmic ray acceleration sites, and presumably have not saturated their magnetic field amplification in this manner. It is even possible that the electron heating itself prevents the magnetic field amplification from
saturating.

We compare the growth rate for waves that heat electrons, $\Gamma\simeq 0.07\omega n_{CR}/n_i$ with the growth rate for magnetic field amplification, $\Gamma=\sqrt{2}k_{\Vert max}v_A$. We find
\begin{equation}
{\delta B^2\over B^2}={\omega _{pi}\over\Omega _{i0}}\sqrt{m_e\over 2m_i}{v_{CR}\over c}{1\over 0.07}\simeq 180\left(v_s\over 5000 {\rm km~s}^{-1}\right)\sqrt{n_i}\left(3\mu G\over B\right)
\left(\gamma _{max}-3\gamma _1/4\over\gamma _{max}-\gamma _1\right)
\end{equation}
The predicted amplified magnetic field is higher than that coming from nonresonant amplification
saturated by resonant scattering, and so that result is unlikely to change. But in this case
the magnetic field never grows to
the level where lower hybrid wave growth competes with magnetic field amplification, so we do not
expect significant electron heating when resonant saturation is important. However the field
in equation 19
is {\em lower} than that expected from nonresonant saturation, and so the electron heating in
a cosmic ray precursor might prevent the full nonlinear stage of that instability from developing, and electron
heating by lower hybrid waves to take over from magnetic field amplification in dissipating the free energy of the
cosmic ray current.
From equation 12, we would also expect it to reduce the depth of the magnetic field shock precursor by a similar factor, i.e. about one order of magnitude.

The level of magnetic field saturation imposed by electron heating on the amplification by
pre-existing turbulence is harder to assess. In the case that the magnetic field is amplified
by the shock itself \citep{giacalone07}, the electron heating should have no effect. In the
case discussed by \citet{beresnyak09} and \citet{drury12}, corresponding to the interaction of the cosmic
ray precursor with pre-shock turbulence, the growth rate is approximately given
by the vorticity resulting in the preshock flow following deceleration by the cosmic ray pressure
gradient. This is $\Gamma\sim v_s\eta\delta\rho /\rho\lambda $, where $\delta\rho /\rho$  expresses
the amplitude of the preshock turbulence, and $\lambda$ is the length scale over which the
density varies. Equating this to the growth rate for lower-hybrid waves yields
$\delta B\sim 2\times 10^7\left(\ln\gamma _{max}-1\right)\delta\rho/\rho\lambda$ G. Specializing to the
forward shock of Cas A, where $\lambda\sim 10^{18}$ cm and $\delta\rho /\rho\sim 10^3$,
$\delta B\sim 2\times 10^{-8}\left(\ln\gamma _{max}-1\right)$ G. This is much lower than the observed field of order $10^{-4}$G, suggesting that such instabilities are not operating. Values of $\lambda$ as low as $10^{15}$ cm would be required. As
\citet{beresnyak09} comment, the reason such instabilities may compete with the nonresonant
current driven instability is that although the growth rate is intrinsically weaker, all the cosmic rays participate in the former. In the
latter, only the very highest energy cosmic rays contribute. But because the growth rate is intrinsically
weaker, the \citet{beresnyak09} instability is more easily saturated by electron heating.

\section{Electron Heating}
\subsection{Electron Heating by Cosmic Rays and the Shock Velocity}
We briefly recap. \citet{ghavamian07} and \citet{vanadelsberg08} report a dependence of
electron temperature to ion temperature, $T_e/T_i$ immediately postshock approximately proportional to $1/v_s^2$
for a selection of supernova remnants exhibiting H$\alpha$ emission.
If $T_i\propto v_s^2$, then $T_e$ is constant. \citet{ghavamian07} put forward an explanation that
electron heating in a cosmic ray precursor of length $L\sim\varkappa /v_s$, where $\varkappa$ is
the cosmic ray diffusion coefficient would lead to such a dependence. The time spent
in the precursor by preshock gas is then $t\sim\varkappa /v_s^2$, leading to electron heating
$E=0.5 mv_e^2 = 0.5 mD_{\Vert\Vert}t$ independent of $v_s$ if the electron velocity diffusion coefficient,
$D_{\Vert\Vert}$, in waves excited in the precursor, is proportional to $v_s^2$. In the specific
case of lower-hybrid waves considered by \citet{ghavamian07}, this is the case. In this discussion, no reference was made to the specific form  of the cosmic ray precursor, but the required
dependence $\varkappa\propto 1/B$ suggests a Bohm-like diffusion coefficient and magnetic
field amplification via the resonant instability.

This far, it has been the purpose of this paper to investigate how far such behavior should
persist into the regime of nonresonant magnetic field amplification. \citet{rakowski08}
invoked magnetic field amplification as a means of ensuring a locally quasi-perpendicular shock
(necessary for the generation of lower hybrid waves), and showed that at high Alfv\'en Mach number, magnetic field amplification has the higher growth rate, while at lower $M_A$ the cosmic ray
generation of lower hybrid waves wins. They estimated a critical $M_A\sim 6v_{inj}/v_s$ which would become $M_A\sim 3v_{inj}/v_s\sim 30$ with the correction to the lower-hybrid wave growth rate
discussed above in equation 18, where $v_{inj}\sim 10v_s$ at a perpendicular shock is the injection velocity for
diffusive shock acceleration. This is very similar to the estimate above in equation 19. In both of these (resonant and nonresonant) cases, the magnetic field amplification and lower hybrid wave
generation is dependent on the streaming velocity of unmagnetized cosmic rays through the
upstream medium.

Figure 1 illustrates schematically how we envisage the electron heating to vary with shock
velocity. At relatively low shock velocities, where the cosmic rays amplify magnetic
field by resonant interactions, $T_e$ is constant, according to the arguments above. At higher
shock velocities, where nonresonant amplification can occur, $T_e$ can increase with shock
velocity, as the precursor lengths over which magnetic field amplification can occur become
larger (equations 5 and 12 for resonant and nonresonant saturation respectively). The curve for resonant
saturation of the nonresonant instability is shown as a dotted line, because in this case magnetic field amplification probably always dominates and quenches the electron heating.

By way of contrast, \citet{matsukiyo10} describes a calculation of instabilities excited by shock
reflected ions at perpendicular shocks, which at low $M_A < 10$ predicts $T_e$
approximately independent of shock velocity, and increasing as $v_s^2$ at higher
$M_A$. This arises because at low $M_A$, the modified two stream instability
grows fastest, generating lower hybrid waves, and is driven to saturation
where the electron gyroradius equals the electron inertial length, $v_{T_e}/\Omega _e\sim c/\omega _{pe}$, giving $n_ek_{\rm B}T_e\sim B^2/8\pi$. Thus so long as
$B^2/n_e$ is constant, shocks of different velocity heat electrons to the same $T_e$. At higher $M_A$, the Buneman instability with faster growth rate takes over,
leading to the sequence of growing Langmuir waves, heating electrons and then ion acoustic waves taking over, as modeled by \citet{cargill88}, leading to $T_e\propto v_s^2$.

\subsection{Effect of Magnetized Cosmic Rays}
We have argued that the shock velocity dependence of the length of shock precursors due to cosmic
ray magnetic field amplification means that higher electron heating should be expected at shocks
where nonresonant magnetic field amplification is dominant over resonant processes. Following
\citet{riquelme10}, another possibility for electron heating at shocks undergoing nonresonant
magnetic field amplification may exist, involving a drift instability of magnetized cosmic rays.

As cosmic rays streaming ahead of a shock amplify magnetic field, more and more cosmic rays at the
lower end of the energy spectrum become magnetized. Their gyroradii in the stronger field are
too small to allow them to stream ahead and contribute to the current that amplifies the field.
Thus, according to the inequality 15, a shock with strong cosmic ray current that saturates
resonantly may transition to nonresonant saturation (and higher magnetic field) if sufficient
cosmic rays become magnetized to reduce $\gamma n_{CR}$ and hence satisfy the inequality.
The resonant saturation of nonresonantly generated field occurs because streaming cosmic rays
are scattered and isotropized by the turbulence. In this case, no electron heating by cosmic rays should occur, because the streaming motion is inhibited. \citet{riquelme10} discuss an
interesting exception to this rule. A drift current associated with magnetized cosmic rays may
also amplify magnetic field. We argue below that it might also generate
lower-hybrid waves and heat electrons.

\citet{riquelme10} give a theory of the Perpendicular Current Driven Instability. In the presence of cross-$B$ density gradients, a cosmic ray drift current may develop along the vector ${\bf \nabla} f_{CR}\times {\bf B}$. \citet{riquelme10} estimate the cosmic ray drift velocity to be $\sim c/2$, so other instabilities besides the magnetic field amplification that they discuss could operate, including the generation of waves (e.g. lower hybrid waves) that could heat electrons.

Therefore, we briefly investigate a reactive instability driven by cosmic ray drift. In perpendicular propagation, the longitudinal part of the cold plasma dielectric tensor is
\begin{equation}
K^L=1+{\omega _{pe}^2\over\Omega _e^2}+{\omega _{pi}^2\over\Omega _i^2-\omega ^2}+{\omega _{pCR}^2\over\Omega _{CR}^2-\omega ^2}{\omega -
{\bf k}\cdot{\bf v}_d\over\omega }=0,
\end{equation}
where $\omega _{pCR}$ and $\Omega _{CR}$ are the cosmic ray plasma
and cyclotron frequencies respectively, and ${\bf v}_d$ is the cosmic ray drift velocity.
We have assumed $\omega << \Omega _e$, the electron gyrofrequency. The dispersion relation is
\begin{equation}
\omega ^5-\omega ^3\Omega _{LH}^2+\omega ^2{\bf k}\cdot{\bf v}_d{n_{CR}\over\left<\gamma\right>n_i}\Omega _{LH}^2+\omega\Omega _{LH}^2\Omega _{CR}^2-{\bf k}\cdot{\bf v}_d{n_{CR}\over\left<\gamma\right>n_i}\Omega _{LH}^2\Omega _i^2=0
\end{equation}
where $\Omega_{LH}^2=\Omega _i\Omega _e$.
As ${\bf k}\cdot {\bf v}_d\rightarrow 0$, the solutions are
\begin{equation}
\omega =0, \quad\omega ^2={\Omega _{LH}^2\over2}\pm{1\over 2}\sqrt{\Omega _{LH}^4-4\Omega_{LH}^2\Omega_{CR}^2}\simeq\Omega _{LH}^2 ~{\rm or}~ \Omega _{CR}^2.
\end{equation}

Reinstating the cosmic ray drift current, we find growing solutions at
$\omega = \pm\Omega _{LH}$ when ${\bf k}\cdot{\bf v}_dn_{CR}/\left<\gamma\right>n_i > 0.39\Omega _{LH}$. With $n_{CR}/\left<\gamma\right>n_i\simeq 3\eta v_s^2/\left<\gamma\right>^2c^2$, $k\simeq\Omega _{LH}/v_i$ where $v_i\sim 10^6$ cm s$^{-1}$ is the ion thermal speed in the precursor, $v_d\simeq c/2$, we find $v_s^2 > 0.26cv_i\left<\gamma\right>^2/\eta$, or
$v_s > 2.8\times 10^8\left<\gamma\right>$ cm s$^{-1}$, assuming $\eta\simeq 0.1$. So depending on the value of $\left<\gamma\right>$ assumed, one might expect to see a contribution to electron heating from magnetized cosmic rays start to become effective at shock velocities of order a few thousand km s$^{-1}$, which should exhibit itself as a break from the $T_e/T_i\propto 1/v_s^2$ law reported by \citet{ghavamian07} and \citet{vanadelsberg08}. Such a contribution also has
to compete with magnetic field amplification by magnetized cosmic rays, which has growth rate \citep{riquelme10}
\begin{equation}
\Gamma _B= 2{J_{CR}\over c}\sqrt{\pi\over\rho}{v_A/c_s\over 1+v_A/c_s} = {\omega _{pi}n_{CR}v_s\over n_ic}{v_A/c_s\over 1+v_A/c_s}.
\end{equation}

The resulting electron temperature may be estimated as follows. The electron velocity will be
\begin{equation}
v_e\simeq {\Omega _{LH}\over k_{\Vert}}\simeq {\Omega _e\over k}
\end{equation}
since lower hybrid waves propagate within a cosine $\sqrt{m_e\over m_i}$ of the perpendicular to
the magnetic field. If we put $k\simeq 0.39\Omega _{LH}\left<\gamma\right>n_i/n_{CR}v_d$ the
electron kinetic energy is
\begin{equation}
{1\over 2}m_ev_e^2 \simeq {m_iv_d^2\over 2\times 0.39^2}{n_{CR}^2\over n_i^2\left<\gamma\right>^2}
\simeq 8\times 10^8 {n_{CR}^2\over n_i^2\left<\gamma\right>^2}
\end{equation}
where we have put $v_d\simeq c/2$. For $n_{CR}/n_i\left<\gamma\right>\sim 10^{-3}$, this gives electron energies of
order 100 - 1000 eV, constant with shock velocity if this ratio is also constant. Requiring
$\Omega _{LH}/v_e < k < \Omega _{LH}/v_i$ for lower hybrid waves yields the following inequality
\begin{equation}
{0.78 v_i\over c} < {n_{CR}\over n_i\left<\gamma\right>} < {0.78 v_e\over c}
\end{equation}
which at $T=10^4$K gives $2\times 10^{-5} < n_{CR}/n_i\left<\gamma\right> < 10^{-3}$, consistent
with the forgoing.

An analytic expression for the lower hybrid wave growth rate can be derived by dropping the last two terms in equation 22, which are of order
$m_e/m_i$ relative to the other through their dependence on $\Omega _{CR}$, and solving the
resulting cubic equation
\begin{equation}
\omega ^3-\omega\Omega _{LH}^2+{\bf k}\cdot{\bf v}_d{n_{CR}\over\left<\gamma\right>n_i}\Omega _{LH}^2=0.
\end{equation}
Using standard procedures \citep{abramowitz84}, the condition for complex roots becomes
${\bf k}\cdot{\bf v}_dn_{CR}/\left<\gamma\right>n_i > \left(2/3\sqrt{3}\right)\Omega _{LH}=0.385
\Omega _{LH}$, in good agreement with the numerical solution above, and with predicted
growth rate $\Gamma _{LH}=2^{-1/3}3^{-1/2}A^{-2/3}\sqrt{A^2/4-1/27}\Omega _{LH}$ where
$A={\bf k}\cdot{\bf v}_dn_{CR}/\left<\gamma\right>n_i\Omega _{LH}$.

In Fig. 2 we plot $\Gamma _{LH}$ against $\log _{10}A$ as a solid line. It increases monatonically from 0 at $A=0.385$, ($\log _{10}A=-0.415$). We compare with $\Gamma _B$ plotted
as dashed lines for values of $c_s/v_A=$ 1, 0.3, 0.1, and 0.03 (from equation 22, in units of $\Omega _{LH}\left<\gamma\right>$). As
the magnetic field becomes stronger, the lower hybrid wave growth rate becomes stronger relative
to the growth rate for magnetic field amplification. At sufficiently large $A$, magnetic field amplification
always wins, but there is always a range of lower $A$ where lower hybrid wave growth dominates.
These dashed curves take a realistic electron-proton mass ratio. \citet{riquelme10} present
simulations with $m_e/m_i = 0.1$, and for this reason we give as dotted curves $\Gamma _B$ for the
reduced mass ratio and the same values of $c_s/v_A$. The magnetic field amplification is much stronger in such
circumstances, and the ``window'' in $A$ where lower hybrid wave growth dominates is much reduced,
consistent with \citet{riquelme10} who do not report any evidence of electron heating in their simulations. \citet{riquelme11} also report a mass ratio dependence in their treatment of electron injection by whistlers.
A detailed assessment of the electron heating at saturation will require
a dielectric tensor accounting for electron thermal motions in place of equation 21, which will
be deferred to a separate paper. However our simple treatment illustrates that in realistic conditions, lower hybrid
wave growth may compete with magnetic field amplification and provide extra electron heating at fast efficient
cosmic ray accelerating shocks.

\section{The Forward Shocks of SN 1006, Cas A, and SNR 0509-67.5}
\subsection{Preamble}
We have argued that electron heating at SNR shocks should break from the behavior discovered in shocks with $v_s<3000$ km s$^{-1}$ by
\citet{ghavamian07} and \citet{vanadelsberg08} at sufficiently high shock velocity where
nonresonant magnetic field amplification sets in. Higher electron temperatures should be expected
in this regime, because the shock precursor over which electron heating becomes longer than in the
purely resonant case, or because magnetized cosmic rays may also contribute to electron heating
through the drift instability treated in subsection 3.3.
As a first step we therefore investigate the electron heating behind the forward shocks of SN 1006 and Cas A, which are faster than those studied in the samples above.

\subsection{SN 1006}
The likely remnant of a Type Ia supernova, SN 1006 is located 500 pc above the galactic plane, expanding into unusually low density interstellar medium. Having sustained high shock speeds over a long period of time, it is the most effective accelerator of cosmic rays of all known SNRs.

SN 1006 was the first SNR from which the X-ray emission was conclusively identified as synchrotron radiation (Koyama et al. 1995, Reynolds 1996). This emission is strong on the NE and SW limbs.
It has been demonstrated that the NE and SW synchrotron limbs represent ``polar caps'' (Willingale et al. 1996; Rothenflug et al. 2004; Reynoso et al. 2013), suggesting that particle acceleration (at least of electrons) is favored in this particular direction, which coincides with the likely direction of the galactic magnetic field (Leckband et al. 1989).
Thus we should tentatively conclude that particle acceleration at quasi-parallel shocks (magnetic field aligned along shock velocity vector) is more favored than at quasi-perpendicular. Cassam-Chena\"i et al. (2008) find the contact discontinuity closer to the forward shock in these regions (NE and SW) than elsewhere in the remnant, though Miceli et al. (2009) reach different conclusions. \citet{rakowski11} study knots of emission apparently ahead of the
blast wave in the SE region. While the regular spacing between the three knots
is consistent with what one would expect from the nonlinear
saturation of the Bell instability, the spectra of these knots indicate that they are most likely ejecta. However in
plasma with an adiabatic index of 5/3, such knots are not expected to move anywhere close to the blast wave, let
alone overtake it. \citet{rakowski11} suggested, following \citet{jun96}, that density perturbations ahead of
the shock advected downstream induce extra vorticity that allows such knots to move ahead of the forward shock. In
the case of SN 1006, situated high above the galactic plane, perturbations induced by a cosmic ray precursor are
a plausible origin of such density structures.

A HESS detection of SN 1006 has been reported \citep{acero10}, consistent with an ambient gas density of 0.05 cm$^{-3}$ \citep{acero07}. Proper motions have been measured in the optical, 0.28'' yr$^{-1}$ (in the NW, see Fig. 1; Winkler et al. 2003), corresponding to a shock velocity of $v_s=2900$ km s$^{-1}$ (assuming a distance of 2.2 kpc), and
in X-rays at 0.48'' yr$^{-1}$ in the NE (Katsuda et al. 2009) giving 5000 km s$^{-1}$, and at 0.3'' - 0.49'' in the NW (Katsuda et al. 2013) indicating higher density in the NW than elsewhere.
Hamilton et al. (2007) derive a reverse shock velocity of 2700 km s$^{-1}$, and determine the expansion velocity of ejecta entering the reverse shock to be 7000 km s$^{-1}$ from HST observations.
The NW limb is believed to have encountered higher densities, ($\sim 0.4$ cm$^{-3}$), including partially neutral material giving rise to optical and UV spectra.

We model SN 1006 as expanding into a uniform density interstellar medium with density 0.05 amu cm$^{-3}$, explosion energy $1\times 10^{51}$ ergs, and ejecta mass 1.4 $M_{\sun}$, yielding
a blast wave velocity of 4700 km s$^{-1}$ and radius 8 pc at a distance of 2.2 kpc \citep{ghavamian02}. Figure 3
shows similar loci of electron temperature and ionization age behind the forward shock for the
cases of no preshock electron heating, and heating to $1\times 10^6$, $3\times 10^6$, and $1\times 10^7$ K respectively moving upwards. We show data points for the SE synchrotron dim region given
by \citet[][including their 95\% confidence limits]{miceli12}.
These indicate a temperature of $5\times 10^6 - 1\times 10^7$ K in the
precursor. This is significantly higher than the electron heating of $3\times 10^6$ K determined
at the NW limb by \citet{ghavamian07}, supporting our hypothesis. Note the this is an electron temperature immediately
postshock, including any heating by adiabatic compression occurring upon shock passage, and so implies
a temperature of $\sim 1\times 10^6$ K in the precursor.

Assuming
otherwise similar cosmic ray acceleration properties at the NW and SW limbs of SN 1006, the presence of neutrals
in the NW, presumably indicative of a region of increased density in the interstellar medium which decelerates
the shock, that must be making the difference. In our view, the shock has been decelerated in the NW to velocities
where resonant magnetic field amplification dominates, whereas in the SE, nonresonant amplification is still
operating, giving rise to the increased electron heating. The direct damping of Alfv\'en waves by
charge exchange reactions is probably insignificant compared to other mechanisms of damping.

\subsection{Cas A}
Cas A offers a complementary case to SN 1006. It is a core collapse supernova remnant, which is expanding into dense remnant stellar wind. Since this ambient medium was presumably photoionized by the radiation from the supernova itself, neutrals are absent from the preshock gas. The preshock magnetic field could well be considerably different
to the ``canonical'' interstellar medium values of $\sim 3 \mu$G. The expansion of the stellar wind from the
stellar atmosphere to the position where it is now encountered by the forward shock of the supernova remnant
dilutes any pre-existing magnetic field to negligible strength. Any preshock magnetic field must be generated by
motions within the stellar wind itself. The equipartition field for a wind moving at 20 km s$^{-1}$ would be
10 $\mu$G, taking a density of 2 amu cm$^{-3}$ from \citet{hwang12}. These parameters correspond to a mass loss
rate of $10^{-4}$ M$_{\sun}$yr$^{-1}$. A lower mass loss rate, or the absence of means for the magnetic field
to reach equipartition would imply a (much) lower preshock magnetic field.

To search for and obtain measured temperatures at the forward shock, we examined thirteen regions at and interior to the outermost filaments of Cas A, as shown in Figure 4.
This search is complicated by two factors: the filaments are mostly dominated by synchrotron emission \citep{gotthelf01,helder08}, and the faint outer regions of the remnant are strongly contaminated by scattering of emission from the bright ejecta ring.

As a dust scattering model is beyond the scope of our work, we chose a conservative approach in subtracting this scattered emission to identify secure examples of thermal X-ray emission associated with the forward shock. We select a local background region in the vicinity of each filament of interest, and model the spectrum in that region including a component for the particle background. The astrophysical component of the background spectrum is generally well-described by a thermal plane parallel shock model ($vpshock$ in XSPEC) with variable abundances characteristic of the ejecta, and fitted temperatures and ionization ages similar to those obtained for ejecta regions by Hwang \& Laming (2012). This background model is then frozen and included in the spectral model for the source region of interest, with only its overall normalization freely fitted. This is a simple and conservative way to remove the effect of the scattered emission from the bright ejecta ring as this scattered emission will vary depending on the relative positions of the source and background region (it is also energy-dependent).  In most cases, the background region is outside the source region, and the fitted normalization is numerically larger, though of comparable scale, to the ratio of geometrical areas on the detector. This is at least consistent with the idea that scattered intensity is higher closer to the ejecta ring, but we found that the scale factor was always fitted higher than the geometrical scale factor, regardless of the relative position of source and background.  Thus the thermal background emission is more likely to be oversubtracted than undersubtracted in our approach, and our identification of thermal emission associated with the forward shock should be conservative.

The source spectrum itself is fitted with a variety of model, including a pure power law and vpshock models with either ejecta-type element abundances or abundances characteristic of Cas A's expected circumstellar environment.  We also considered models combining a power law with a forward shock type model. Regions were rejected for the purpose of this study if they were better fitted with ejecta-type abundances, or a power law or if the best-fitting thermal model for the forward shock had ionization age consistent with zero within its 90\% confidence errors.  The actual fitted ionization ages cannot be taken too seriously, as the spectral models are not likely to be very accurate at these very low ionization ages.  The most that can be gleaned is that the thermal emission is present and that that ionization age is low.
Possibly some of the rejected regions could actually contain detectable emission from the forward shock, but this could not be confidently determined without a sophisticated and careful treatment of the scattered background. Having been chosen conservatively, the two regions that remain are likely to represent instances of thermal emission associated with the forward shock. The best fitting of the models that we examined for these two regions were a combined power law and thermal forward shock model. Their fitted temperatures and ionization ages are given in Table 1, with 90\% confidence limits.

We model Cas A very similarly to the preferred model for Cas A in \citet{hwang12}; 3 $M_{\sun}$ ejecta, $2.4\times 10^{51}$ ergs explosion energy, a product of blast wave radius and circumstellar density
$\rho r_f^2 = 16$ amu pc$^2$, and the radius of a circumstellar bubble $R_{bub}=0.3$ pc. The blast
wave has velocity 5000 km s$^{-1}$ (at the current epoch) and is at a radius of 2.6 pc. The preshock medium
is taken to be 49\% H, 49\% He, and 2\% N by mass, reflecting the enrichment of the outer stellar layers in N by the
CNO process. Relative to H, He and N are enhanced over solar values by factors of approximately 2 and 20 respectively.

Figure 5 shows the locus of electron temperature and ionization age (the product of electron density and time, $n_et$) behind the forward shock for different values of
electron heating in the precursor. From the lowest to the highest, they represent no heating, and
heating to temperatures of $1\times 10^6$, $3\times 10^6$, $1\times 10^7$ and $3\times 10^7$ K
respectively. Electrons are further heated by adiabatic compression on passage through the shock.
The two points labeled ``a'' and ``b'' come from fits in \citet{vink03}, one from a synchrotron bright portion of the blast wave with higher electron temperature, and one from a synchrotron dim
region with lower electron temperature. Also shown on Fig. 5 are points from fits to regions labeled ``s4'' and ``sw'', with locations within Cas A shown in Fig. 4. These both have lower $n_et$ than regions ``a'' and ``b'',
and so have undergone passage through the forward shock more recently. They also broadly support the inference from
the other two, that of a precursor electron temperature in the range $1\times 10^7 = 3\times 10^7$ K, again
significantly higher than the corresponding value from an extrapolation of the survey of \citet{ghavamian07}. Considering that the
preshock magnetic field may be lower, possibly considerably lower, than that in SN 1006, the inference of
postshock magnetic fields of order 100 $\mu$G \citep{vink03} suggests that nonnresonant magnetic field amplification
and saturation should be at work, if cosmic rays are responsible. Unfortunately, it does not appear possible to
detect such a precursor observationally, because of scattered light from the bright remnant interior. We also
note that the circumstellar medium of Cas A, unlike that presumed for SN 1006, is naturally clumpy, and that in
such circumstances, magnetic field amplification by the interaction of the shock with such pre-existing turbulence
\citep{giacalone07}, or variations upon this mechanism, \citep[e.g.][]{beresnyak09,drury12} cannot so easily be ruled out.

\subsection{SNR 0509-67.5} In our focus on higher shock velocities than those considered by \citet{ghavamian07},
results derived from SN 1006 and Cas A, while supportive of our hypothesis, are limited by the fact that the emission
we observe comes from a region downstream of the shock itself, and electron heating at the shock front is derived
by model extrapolation. This is presumably because the H$\alpha$ emission studied by \citet{ghavamian07} is
strongest in regions where the shock encounters denser regions of interstellar medium; sufficiently dense that
neutrals can survive against ionization, and the shock naturally decelerates in such regions.
One possible example of a faster shock displaying H$\alpha$ emission is
SNR 0509-67.5 \citep{helder10}. The remnant has forward shock velocities in the range 5200 - 6300 km s$^{-1}$ \citep{ghavamian07b}, derived from the width of the Ly$\beta$ line.

The H$\alpha$ line profile consists of two components; a narrow feature arising as
preshock neutrals diffuse through the shock and are excited before being ionized by shocked ions and electrons, and
a broad component with an origin in charge exchange, as a preshock neutral has its electron captured by a shocked proton. The intensity ratio between these two can be a diagnostic of the electron temperature. If the neutrals are
primarily destroyed by electron impacts, then the broad/narrow intensity ratio will be small (i.e. not enough slow neutrals survive long enough to become fast neutrals by charge exchange), but if the electrons are inefficient at
ionizing neutrals, because of insignificant electron heating at the shock, the broad component becomes comparable in
intensity to the narrow feature.

In SNR 0509-67.5, the broad component intensity is significantly smaller than that of the narrow component, indicating the presence of electron heating. The analysis of \citet{helder10} also accounts for shock energy losses
to cosmic rays, corroborating the shock speeds derived by \citet{ghavamian07b}, and finding electron/proton
temperature ratios in the range 0.2 - 1, significantly larger than those found by \citet{ghavamian07}. While
more uncertain than the SN 1006 or Cas A results given above, the elevated electron temperature is determined at the
forward shock itself, and further supports the argument of this paper.

\section{Conclusions}
In this paper we have attempted to hypothesize how electron heating at collisionless shocks
might behave in the  absence of neutrals in the preshock medium, and in the presence of
magnetic field amplification by the nonresonant cosmic ray current driven instability.
Compared to shocks dominated by resonant magnetic field amplification where the prescursor
length $L\simeq\varkappa /v_s$, shocks with magnetic field amplified by the nonresonant
instability have $L$ decreasing less quickly with increasing $v_s$. The time spent in
the precursor, $t=L/v_s$ decreases more slowly than $1/v_s^2$, leading to an increase in
the electron temperature expected if the electron heating is mediated by cosmic ray generated
lower hybrid waves with velocity diffusion coefficient proportional to $v_s^2$. Thus at
shocks of higher velocities than those in the samples of \citet{ghavamian07} and \citet{vanadelsberg08}, we speculate that the electron temperature should rise with increasing shock speed.
Analysis of the electron temperature behind higher speed shocks in Cas A and in SN 1006
supports this view, within observational uncertainties, as do observations of Balmer emission
from SNR 0509-67.5 \citep{helder10}.

At still higher shock speeds, a drift
instability due to magnetized cosmic rays may also begin to heat electrons. The threshold
speed where this might begin is uncertain, but of order 10,000 km s$^{-1}$. Observations of
higher velocity shocks are obviously very interesting in this regard. Of accessible supernova
remnants, 1E0102 and SN 1987A would seem to be the prime candidates. \citet{hughes00} infer a forward shock velocity
for 1E0102 from proper motion measurements of 6000 km s$^{-1}$. This forward shock, however, is running into a
remnant stellar wind, which may itself have a speed of 2000 - 3000 km s$^{-1}$, thus reducing the actual forward
shock velocity with respect to the upstream medium. Further, \citet{flanagan04} only infer expansion velocities
of order 1000 km s$^{-1}$ from observations of the SNR ejecta. The forward shock of SN 1987A is currently running
into the inner circumstellar ring \citep{zhekov05}. The fast deceleration means that the shock velocity is uncertain, and the spectrum obtained necessarily has contributions from decelerated and relatively undecelerated portions of the blast wave, making precise interpretation difficult. A further complication here is that SN 1987A is relatively young, and it is unclear how much energy has yet accumulated in cosmic rays at its forward shock.

\acknowledgements This work was supported by the NASA ADP program (UH, JML and CER), and by basic
research funds of the Office of Naval Research (JML and CER). JML would also like to acknowledge the hospitality of R. Bingham during a visit to the Rutherford Appleton Laboratory, where some of this work was started.

\appendix
\section{Cosmic Ray Distributions}
The cosmic ray distribution is taken to be \begin{equation}
f_{CR}\left(p\right)={n_{CR}p_0^{1+a}p_{max}^{1+a}\over 4\pi\left(p_{max}^{1+a}-p_0^{1+a}\right) p^{4+a}}
\end{equation}
where $p_0$ and $p_{max}$ are the lower and upper limits on the cosmic ray momentum respectively. The cosmic ray kinetic energy is
\begin{eqnarray}
 & & E_{CR}\simeq\int _{p_0}^{mc}f_{CR}{p^2\over 2m}4\pi p^2dp +\int _{mc}^{p_{max}}
f_{CR}\left(pc-mc^2\right) 4\pi p^2dp\nonumber\\
& & \simeq n_{CR}{p_0^2\over 2m}{\left[1-\left(mc/p_0\right)^{1-a}\over a-1\right]}
+n_{CR}p_0c\left(p_0\over mc\right)^a\left[1-\gamma_{max}^{-a}\over a\right]
\end{eqnarray}
where $\gamma _{max}=p_{max}/mc$ and $p_{max} >> p_0$. When $a=1$ or $a=0$ in the first or second terms respectively, the quantities in square brackets become $\ln\left(mc/p_0\right) -1$ and
$\ln\gamma _{max}-1$ respectively.
The relativistic contribution dominates, except where $a >> 1$ and $mc >> p_0$. Given that $a=0$ reproduces the standard cosmic ray spectrum resulting from first order Fermi acceleration at a strong shock, and the $p_0\sim mc$ is frequently required for injection, we neglect the nonrelativistic contribution in the following. Specializing to $a=0$ for the time being
we also write the cosmic ray number density $n_{CR}$ in terms of the ratio of cosmic ray pressure to shock ram pressure $\eta= P_{CR}/\rho v_s^2$ as follows
\begin{equation}
n_{CR}={E_{CR}\over p_0c\left(\ln\gamma _{max}-1\right)}={3\eta\rho v_s^2\over p_0c\left(\ln\gamma _{max}-1\right)}.
\end{equation}
With $\rho = n_im$ and $p_0=mv_{inj}$ this simplifies to
\begin{equation}
{n_{CR}\over n_i}={3\eta v_s^2\over cv_{inj}\left(\ln\gamma _{max}-1\right)}\simeq
{3\eta v_s^2\over c^2\left(\ln\gamma _{max}-1\right)}.
\end{equation}
The average cosmic ray kinetic energy
\begin{equation}
\left<E_{CR}\right>={E_{CR}\over n_{CR}}=p_0c\left(\ln\gamma _{max}-1\right) =\left(\left<\gamma\right>-1\right)mc^2
\end{equation}
so $\left<\gamma\right>\simeq \left(\ln\gamma _{max}-1\right)\times v_{inj}/c +1\simeq\ln\gamma _{max}$ if $v_{inj}\simeq c$.

Considering a subpopulation of unmagnetized cosmic rays with $p > p_1$, their density $n_{CR}^{\prime}$ is given by
\begin{equation} n_{CR}^{\prime}= \int _{p_1}^{p_{max}}f_{CR}4\pi p^2dp =
n_{CR}\left(\gamma _{max}-\gamma _1\right)/\gamma _{max}\gamma _1,
\end{equation}
and the average unmagnetized cosmic ray kinetic energy, $E_{CR}^{\prime}$ by
\begin{equation} E_{CR}^{\prime}= \int _{p_1}^{p_{max}}f_{CR}\left(pc-mc^2\right)4\pi p^2dp =
n_{CR}mc^2\left(\ln\gamma _{max}-\ln\gamma _1 +1/\gamma _{max} -1/\gamma _1\right).
\end{equation}
These lead to the average Lorentz factor for the unmagnetized cosmic rays of
\begin{equation} \left<\gamma ^{\prime}\right> = {\gamma _1\gamma _{max}\over \gamma _{max}
-\gamma _1}\ln {\gamma _{max}\over\gamma _1}.
\end{equation}

For $a\ne 0$, we give the following more general results,
\begin{eqnarray}
& & {n_{CR}\over n_i}={3\eta v_s^2\over c^2}\left(c\over v_{inj}\right)^a{a\over 1-\gamma _{max}^{-a}}\\
& & \left<\gamma\right>=1+\left(v_{inj}\over c\right)^{1+a}\left[1-\gamma _{max}^{-a}\over a\right]\\
& & n_{CR}^{\prime}={n_{CR}\over 1+a}{\gamma _{max}^{1+a}-\gamma _1^{1+a}\over\gamma _{max}^{1+a}
\gamma _1^{1+a}}\left(v_{inj}\over c\right)^{1+a}\\
& & \left<\gamma ^{\prime}\right>=\left(v_{inj}\over c\right)^{1+a}{\gamma _{max}\gamma _1\over a}
{\gamma _{max}^a-\gamma _1^a\over\gamma _{max}^{1+a}-\gamma _1^{1+a}}.\\
\end{eqnarray}

\begin{table}
\caption{Cas A Forward Shock Fit Results}
\label{table1}
\centering
\begin{tabular}{lrrr}
region & $kT$ (keV) & $n_et$ (cm$^{-3}$s$^{-1}$) & $\chi ^2$/dof\\
\hline
sw & 2.6 [2.2 - 3.0]& 8.4e9 [7.4e9 - 9.3e9]& 1.48\\
s4 & 2.6 [2.3 - 2.9]& 9.6e8 [7.4e8 - 1.3e9]& 1.21\\

\hline
\end{tabular}
\end{table}

\clearpage
\begin{figure}
\centerline{\includegraphics[scale=1.0]{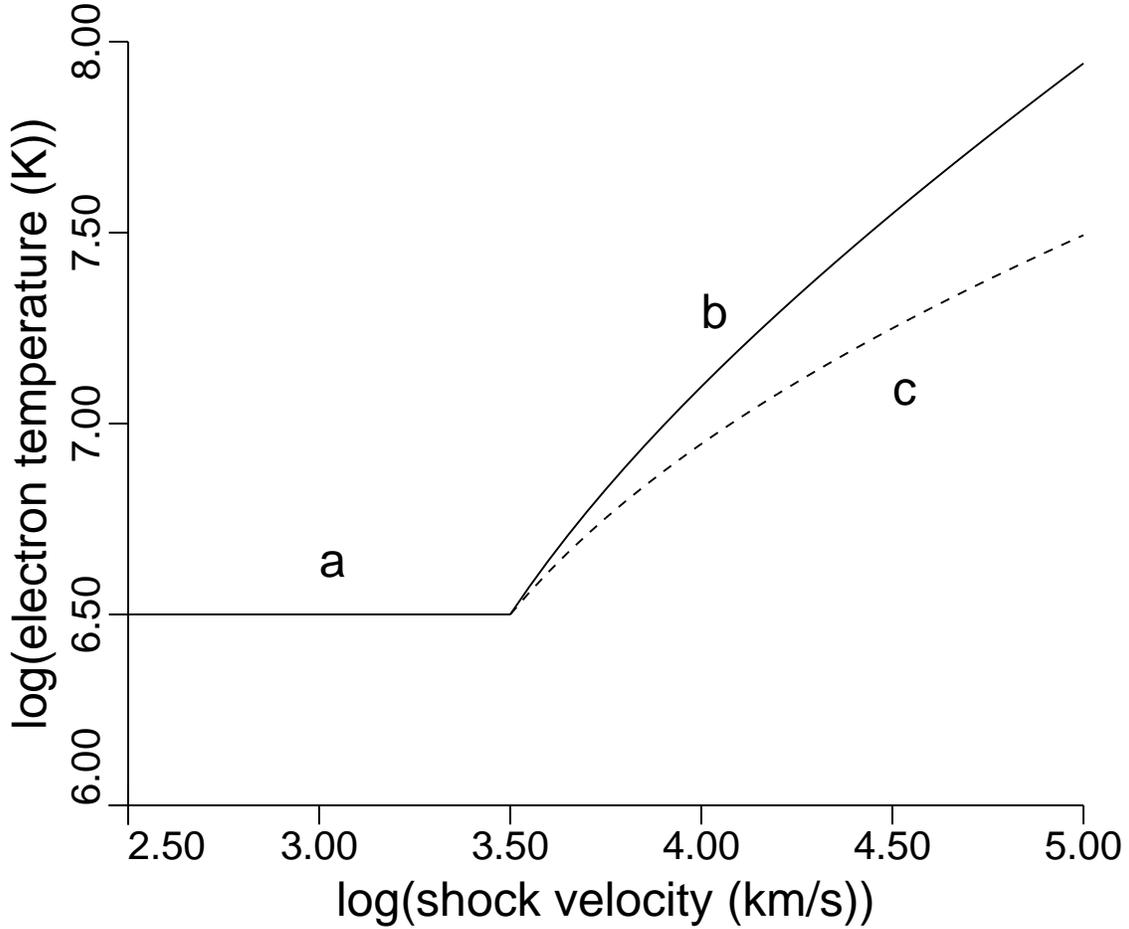}}
\caption{Schematic diagram showing the hypothesized variation of post shock electron temperature
with shock velocity. At low shock speeds, the branch of the plot labeled ``a'' shows constant
electron heating where cosmic rays resonantly amplify magnetic field. With constant cosmic ray
diffusion coefficient $D$, the length of the precursor $L\sim \varkappa/v_s$, so the time spent in the
precursor by preshock gas $t\sim \varkappa/v_s^2$, leading to electron heating $\sim D_{\Vert\Vert} t$ independent
of $v_s$ where $D_{\Vert\Vert}\propto v_s^2$ is the diffusion coefficient in velocity space for electrons
in lower hybrid waves. The branch labeled ``c'' shows the behavior when magnetic field in
nonresonantly amplified, but resonantly saturated and $L\propto v_s^{-4/5}\ln v_s$, and branch
``b'' shows the completely nonresonant case, with $L\propto v_s^{-1/2}\ln v_s$. The break
between resonant and nonresonant amplification at $v_s=3000$ km s$^{-1}$ is motivated by the
discussion in \citet{marcowith10}, but remains uncertain. Branch ``c'' is shown as a dashed line, because in this
case the magnetic field amplification growth rate is always larger than that for lower hybrid waves.}
\end{figure}

\begin{figure}
\centerline{\includegraphics[scale=1.0]{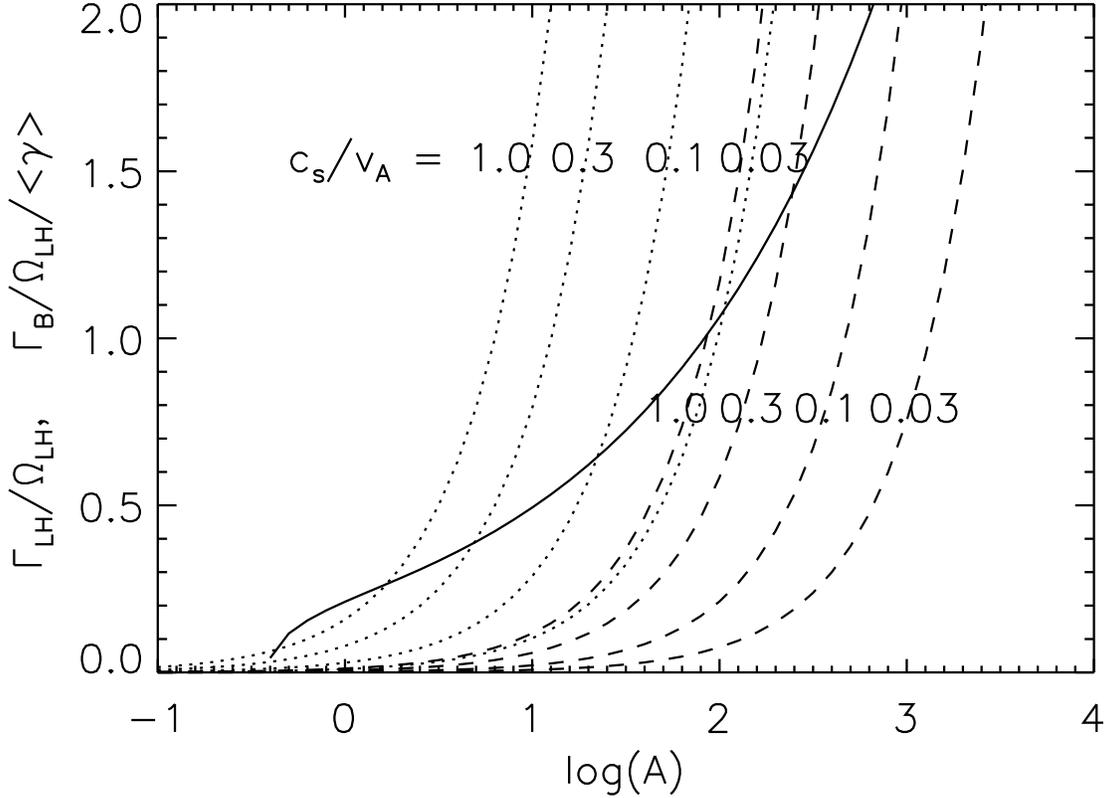}}
\caption{Growth rates for lower hybrid waves ($\Gamma _{LH}$) and magnetic field
amplification ($\Gamma _B$), scaled by $\Omega _{LH}$, by magnetized cosmic rays in a shock precursor, against
$\log _{10}A=\log _{10}\left({\bf k}\cdot{\bf v}_dn_{CR}/\left<\gamma\right>n_i\Omega _{LH}\right)$.
The solid curve gives $\Gamma _{LH}$. Dashed curves
give values of $\Gamma _B$ for $c_s/v_A =$ 1, 0.3, 0.1, 0.03 successively from the left, with
higher growth for lower $c_s/v_A$. At sufficiently high $A$, $\Gamma _B$ is always dominant, but a region of lower
$A$ exists where $\Gamma _{LH}$ is larger, and lower-hybrid wave growth may occur.
The dotted curves give similar values for $\Gamma _B$, but using an
electron-proton mass ratio of 10, to match this parameter in Riquelme \&
Spitkovsky (2010). Lower hybrid waves, and the ensuing electron heating, are much less favored in such conditions,
because the magnetic field amplification is stronger.\label{fig2}}
\end{figure}

\begin{figure}
\centerline{\includegraphics[scale=1.0]{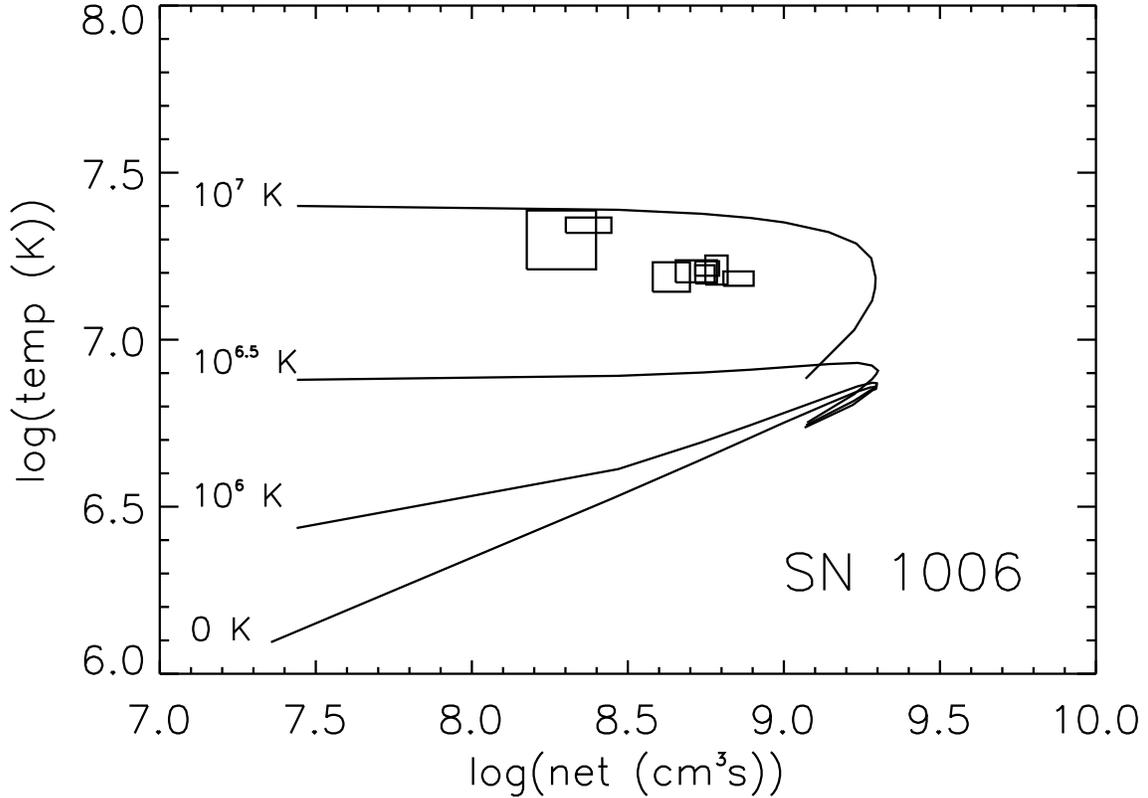}}
\caption{The locus of electron temperature $T_e$ against ionization age $n_et$ for the
forward shock of SN 1006, for four different degrees of electron-ion equilibration at
the shock. The model is for a $1\times 10^{51}$ erg explosion, 1.4$M_{\sun}$ ejecta, interstellar medium density of 0.05 cm$^{-3}$, and an ejecta envelope power law of 7.
The lowest curve shows the case of no equilibration. The succeeding curves give
models for electron heating in the shock precursor to temperatures of $1\times 10^6$, $3\times 10^6$ and $1\times 10^7$ K. On passage through the shock these temperatures increase by a
further factor of $4^{2/3}$ due to adiabatic compression. The boxes indicate data points
from \citet{miceli12}, with 95\% confidence limits.
\label{fig4}}
\end{figure}

\begin{figure}
\centerline{\includegraphics[scale=1.0]{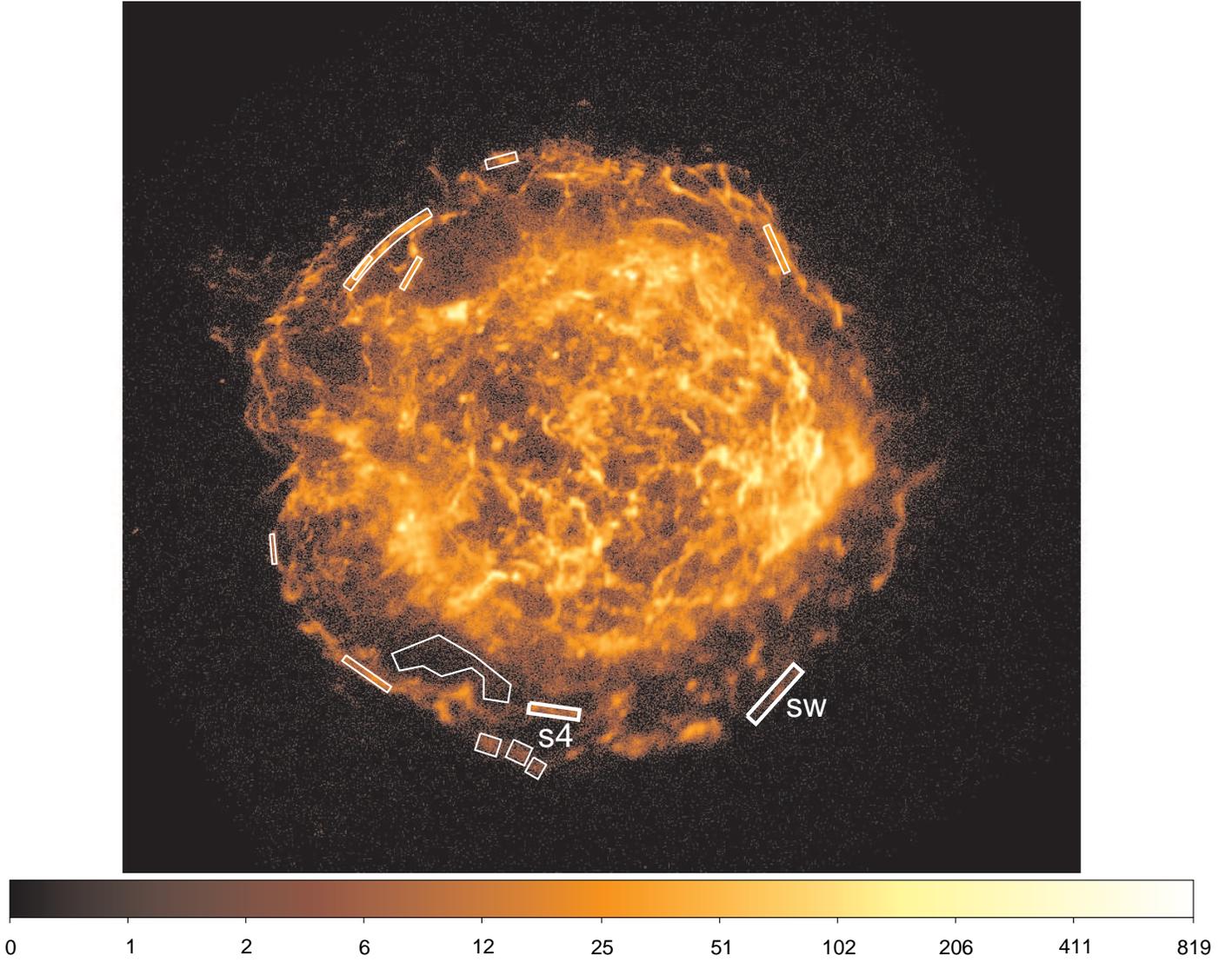}}
\caption{Positions of the thirteen regions in the Cas A SNR examined for this paper are shown superposed on a log intensity image of the 4-6 keV X-ray continuum of Cas A from Hwang et al. 2004, and this work.  The "s4" and "sw"
regions are labelled.
\label{fig5}}
\end{figure}

\begin{figure}
\centerline{\includegraphics[scale=1.0]{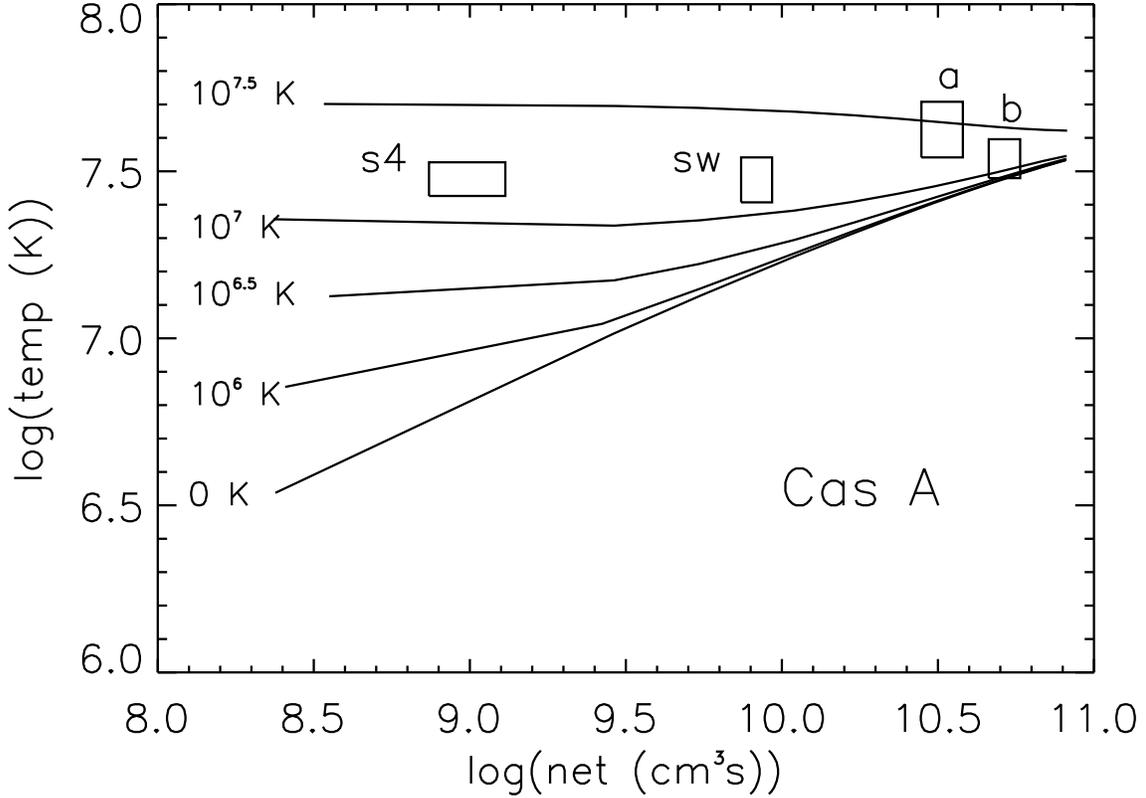}}
\caption{The locus of electron temperature $T_e$ against ionization age $n_et$ for the
forward shock of Cassiopeia A, for four different degrees of electron-ion equilibration at
the shock. The lowest curve shows the case of no equilibration. The succeeding curves give
models for electron heating in the shock precursor to temperatures of $1\times 10^6$, $3\times 10^6$,
$1\times 10^7$ and $3\times 10^7$ K. On passage through the shock these temperatures increase by a
further factor of $4^{2/3}$ due to adiabatic compression. The boxes indicate data points
from \citet{vink03}; ``a'' comes from a bright X-ray synchrotron filament, while ``b'' comes
from a dim synchrotron filament.
\label{fig3}}
\end{figure}


\begin{thebibliography}{}
\bibitem[Abramowitz \& Stegun(1984)]{abramowitz84}Abramowitz, M., \& Stegun, I. A. 1984, Handbook of Mathematical Functions, (Verlag Harri Deutsch: Frankfurt am Main)
\bibitem[Acero et al.(2007)]{acero07}Acero, F., Ballet, J. \& Decourchelle, A. 2007, \aap, 475, 883
\bibitem[Acero et al.(2010)]{acero10}Acero, F., Aharonian, F., Akhperjanian, A. G., et al. 2010,  \aap, 516, A62
\bibitem[Bamert et al.(2004)]{bamert04}Bamert, K., Kallenbach, R., Ness, N. F., et al. 2004, \apj, 601, L99
\bibitem[Bell \& Lucek(2001)]{bell01}Bell, A. R., \& Lucek, S. G. 2001,
 \mnras, 321, 433
\bibitem[Bell(2004)]{bell04}Bell, A. R. 2004, \mnras, 353, 550
\bibitem[Bell(2005)]{bell05}Bell, A. R. 2005, \mnras, 358, 181
\bibitem[Bell et al.(2013)]{bell13}Bell, A. R., Schure, K. M., Reville, B., \& Giacinti, G. 2013, \mnras, 431, 415
\bibitem[Beresnyak et al.(2009)]{beresnyak09}Beresnyak, A., Jones, T. W., \& Lazarian, A. 2009, \apj, 707, 1541
\bibitem[Caprioli \& Spitkovsky(2012)]{caprioli13}Caprioli, D., \& Spitkovsky, A. 2013, \apj, 765, L20
\bibitem[Cargill \& Papadopoulos(1988)]{cargill88}Cargill, P. J., \& Papadopoulos, K. 1988, \apj, 329, L29
\bibitem[Cassam-Chena\"i et al.(2007)]{chenai07}Cassam-Chena\"i, G., Hughes, J. P., Ballet, J., \& Decourchelle, A. 2007, \apj, 665, 315
\bibitem[Cassam-Chena\"i et al.(2008)]{chenai08}Cassam-Chena\"i, G., Hughes, J. P., Reynoso, E. M., Badenes, C. \& Moffett, D. 2008, \apj, 680, 1180
\bibitem[Drury \& Downes(2012)]{drury12}Drury, L. O'C., \& Downes, T. P. 2012, \mnras, 427, 2308
\bibitem[Flanagan et al.(2004)]{flanagan04}Flanagan, K. A., Canizares, C. R., Dewey, D., Houck, J. C., Fredericks, A. C., Schattenburg, M. L., Markert, T. H., \& Davis, D. S. 2004, \apj, 605, 230
\bibitem[Gargat\'e et al.(2010)]{gargate10}Gargat\'e, L., Fonseca, R. A., Niemiec, J., Pohl, M., Bingham, R., \& Silva, L. O. 2010, ApJ, 711, L127
\bibitem[Gargat\'e \& Spitkovsky(2012)]{gargate12}Gargat\'e, L., \& Spitkovsky, A. 2012, ApJ, 744, 67
\bibitem[Ghavamian et al.(2002)]{ghavamian02}Ghavamian, P., Winkler, P. F., Raymond, J. C., \& Long, K. S. 2002, \apj, 572, 888
\bibitem[Ghavamian et al.(2007a)]{ghavamian07}Ghavamian, P., Laming, J. M., \& Rakowski, C. E. 2007, \apj, 654, L69
\bibitem[Ghavamian et al.(2007b)]{ghavamian07b}Ghavamian, P., Blair, W. P., Sankrit, R., Raymond, J. C., \& Hughes, J. P. 2007, \apj, 664, 304
\bibitem[Ghavamian et al.(2013)]{ghavamian13}Ghavamian, P., Schwartz, S. J., Mitchell, J., Masters, A., \& Laming, J. M. 2013, Space Science Reviews, 178, 633
\bibitem[Giacalone \& Jokipii(2007)]{giacalone07}Giacalone, J., \& Jokipii, J. R. 2007, \apj, 663, L41
\bibitem[Gotthelf et al.(2001)]{gotthelf01}Gotthelf, E. V., Koralesky, B., Rudnick, L., Jones, T. W., Hwang, U., \& Petre, R. 2001 \apj, 552, L39
\bibitem[Hamilton et al.(2007)]{hamilton07}Hamilton, A. J. S., Fesen, R. A., \& Blair, W. P. 2007, \mnras, 381, 771
\bibitem[Helder et al.(2010)]{helder10}Helder, E. A., Kosenko, D. \& Vink, J. 2010, \apj, 719, L140
\bibitem[Helder \& Vink(2008)]{helder08}Helder, E. A., \& Vink, J. 2008, \apj, 686, 1094
\bibitem[Hughes et al.(2000)]{hughes00}Hughes, J. P., Rakowski, C. E., \& Decourchelle, A. 2000, \apj, 543, L61
\bibitem[Jun et al.(1996)]{jun96}Jun, B.-I., Jones, T. W., \& Norman, M. L. 1996, \apj, 468, L59
\bibitem[Hwang \& Laming(2012)]{hwang12}Hwang, U., \& Laming, J. M. 2012, \apj, 746, 130
\bibitem[Katsuda et al.(2009)]{katsuda09}Katsuda, S., Petre, R., Long, K. S., Reynolds, S. P., Winkler, P. F., Mori, K., \& Tsunemi, H. 2009,\apj, 692, L105
\bibitem[Katsuda et al.(2013)]{katsuda13}Katsuda, S., Long, K. S., Petre, R., Reynolds, S. P., Williams, B. J.,
\& Winkler, P. F. 2013, \apj, 763, 85
\bibitem[Koyama et al.(1995)]{koyama95}Koyama, K., Petre, R., Gotthelf, E. V., Hwang, U., Matsuura, M., Ozaki, M., \& Holt, S. S. 1995, Nature, 378, 255
\bibitem[Laming(2001a)]{laming01a}Laming, J. M. 2001a, \apj, 546, 1149
\bibitem[Laming(2001b)]{laming01b}Laming, J. M. 2001b, \apj, 563, 828
\bibitem[Laming et al.(2013)]{laming13}Laming, J. M., Moses J. D., Ko, Y.-K., Murphy, R. J., Ng, C. K., Rakowski, C. E., \& Tylka, A. J. 2013, ApJ, 770, 73
\bibitem[Leckband et al.(1989)]{leckband89}Leckband, J. A., Spangler, S. R., \& Cairns, I. H. 1989, \apj, 338, 963
\bibitem[Long et al.(2003)]{long03}Long, K. S., Reynolds, S. P., Raymond, J. C., Winkler, P. F., Dyer, K. K., \& Petre, P. 2003, \apj, 586, 1162
\bibitem[Lucek \& Bell(2000)]{lucek00}Lucek, S. G., \& Bell, A. R. 2000, \mnras, 314, 65
\bibitem[Luo \& Melrose(2009)]{luo09}Luo, Q., \& Melrose, D. 2009, \mnras, 397, 1402
\bibitem[Marcowith \& Casse(2010)]{marcowith10}Marcowith, A., \& Casse, F. 2010, \aap, 515, 90
\bibitem[Matsukiyo(2010)]{matsukiyo10}Matsukiyo, S. 2010, Physics of Plasmas, 17, 042901
\bibitem[Miceli et al.(2009)]{miceli09}Miceli, M., Bocchino, F., Iakubovskyi, D., Orlando, S., Telezhinsky, I., Kirsch, M. G. F., Petruk, O., Dubner, G., \& Castelletti, G. 2009, \aap, 501, 239
\bibitem[Miceli et al.(2012)]{miceli12}Miceli, M., Bocchino, F., Decourchelle, A., Maurin, G., Vink, J., Orlando, S., Reale, F., \& Broersen, S. 2012, \aap, 546, A66
\bibitem[Morlino et al.(2010)]{morlino10}Morlino, G., Amato, E., Blasi, P., \& Caprioli, D. 2010, \mnras, 405, L21
\bibitem[Nekrasov \& Shadmehri(2012)]{nekrasov12}Nekrasov, A. K., \& Shadmehri, M. 2012, \apj, 756, 77
\bibitem[Niemiec et al.(2010)]{niemiec10}Niemiec, J., Pohl, M., Bret, A., \& Stroman, T. 2010, \apj, 709, 1148
\bibitem[Rakowski et al.(2008)]{rakowski08}Rakowski, C. E., Laming, J. M., \& Ghavamian, P. 2008, \apj, 684, 348
\bibitem[Rakowski et al.(2009)]{rakowski09}Rakowski, C. E., Ghavamian, P., \& Laming, J. M. 2009, \apj, 696, 2195
\bibitem[Rakowski et al.(2011)]{rakowski11}Rakowski, C. E., Laming, J. M., Hwang, U., Eriksen, K. A., Ghavamian, P., \& Hughes, J. P. 2011, \apj, 735, L21
\bibitem[Reville \& Bell(2013)]{reville13}Reville, B., \& Bell, A. R. 2013, \mnras, 430, 2873
\bibitem[Reynolds(1996)]{reynolds96}Reynolds, S. P. 1996, \apj, 459, L13
\bibitem[Reynoso et al.(2013)]{reynoso13}Reynoso, E. M., Hughes, J. P., \& Moffett, D. A. 2013, \aj, 145, 104
\bibitem[Riquelme \& Spitkovsky(2009)]{riquelme09}Riquelme, M. A., \& Spitkovsky, A. 2009, \apj, 694, 626
\bibitem[Riquelme \& Spitkovsky(2010)]{riquelme10}Riquelme, M. A., \& Spitkovsky, A. 2010, \apj, 717, 1054
\bibitem[Riquelme \& Spitkovsky(2011)]{riquelme11}Riquelme, M. A., \& Spitkovsky, A. 2011, \apj, 733, 63
\bibitem[Rothenflug et al.(2004)]{rothenflug04}Rothenflug, R., Ballet, J., Dubner, G., Giacani, E., Decourchelle, A., \& Ferrando, P. 2004, \aap, 425, 121
\bibitem[Stroman et al.(2009)]{stroman09} Stroman, T., Pohl, M., \& Niemiec, J. 2009, \apj, 706, 38
\bibitem[Stroman et al.(2012)]{stroman12}Stroman, T., Pohl, M., Niemiec, J., \& Bret, A. 2012, \apj, 746, 24
\bibitem[van Adelsberg et al.(2008)]{vanadelsberg08}van Adelsberg, M., Heng, K., McCray, R., \& Raymond, J. C. 2008, \apj, 689, 1089
\bibitem[Vink \& Laming(2003)]{vink03}Vink, J., \& Laming, J. M. 2003, \apj, 584, 758
\bibitem[V\"olk et al.(2005)]{volk05}V\"olk, H. J., Berezhko, E. G., \&
Ksenofontov, L. T. 2005, \aap, 433, 229
\bibitem[Warren et al.(2005)]{warren05}Warren, J. S., Hughes, J. P., Badenes, C., Ghavamian, P., McKee, C. F., Moffett, D., Plucinsky, P. P., Rakowski, C., Reynoso, E., \& Slane, P.  2005, \apj, 634, 376
\bibitem[Wiener et al.(2013)]{wiener13}Wiener, J., Zweibel, E. G., \& Peng Oh, S. 2013, arXiv:1301.4445
\bibitem[Willingale et al.(1996)]{willingale96}Willingale, R., West, R. G., Pye, J. P., \& Stewart, G. C. 1996, \mnras, 278, 749
\bibitem[Winkler et al.(2003)]{winkler03}Winkler, P. F., Gupta, G., \& Long, K. S. 2003, \apj, 585, 324
\bibitem[Winkler et al.(2014)]{winkler14}Winkler, P. F., Williams, B. J., Reynolds, S. P., Petre, R., Long, K. S., Katsuda, S., \& Hwang, U. 2014, \apj, 781, 65
\bibitem[Winske \& Leroy(1984)]{winske84}Winske, D., \& Leroy, M. M. 1984, J. Geophys. Res., 89, 2673
\bibitem[Yamazaki et al.(2004)]{yamazaki04}Yamazaki, R., Yoshida, T., Terasawa, T., Bamba, A., \& Koyama, K. 2004, \aap, 416, 595
\bibitem[Zank et al.(2006)]{zank06}Zank, G. P., Li, G., Florinski, V., Hu, Q., Lario, D., \& Smith, C. W. 2006, J. Geophys. Res. Space Phys., 111, 6108
\bibitem[Zhekov et al.(2005)]{zhekov05}Zhekov, S. A., McCray, R., Borkowski, K. J., Burrows, D. N., \& Park, S. 2005, \apj, 628, L127
\end{thebibliography}
\end{document}